\documentclass[prd,preprint,showpacs]{revtex4}
\usepackage{epsfig,chay}
\begin{document}
\title{Rare radiative exclusive $B$ decays  in
  soft-collinear effective theory}
\author{Junegone Chay}
\email{chay@korea.ac.kr}
\author{Chul Kim}
\affiliation{Department of Physics, Korea University, Seoul 136-701,
Korea}
\preprint{KUPT--03--03} 
\begin{abstract}
We consider rare radiative $B$ decays such as
$\overline{B}\rightarrow K^* \gamma$ or $\overline{B} \rightarrow \rho
\gamma$ in soft-collinear effective theory, and show that the decay
amplitudes are factorized to all orders in $\alpha_s$ and at leading
order in $\Lambda_{\mathrm{QCD}}/m_b$. 
By employing two-step matching, we classify the operators for
radiative $B$ decays in powers of a small parameter $\lambda(\sim
\sqrt{\Lambda_{\mathrm{QCD}}/m_b})$ and obtain the relevant operators
to order $\lambda$ in
$\mathrm{SCET}_{\mathrm{I}}$. These operators are constructed with or
without spectator quarks including the four-quark operators
contributing to annihilation and $W$-exchange
channels. And we employ $\mathrm{SCET}_{\mathrm{II}}$ where the small
parameter becomes of order $\Lambda_{\mathrm{QCD}}/m_b$, and evolve the
operators in order to compute the decay amplitudes for rare radiative
decays in soft-collinear effective theory. We show
explictly that the contributions from the annihilation channels and
the $W$-exchange channels vanish at leading order in SCET. We present
the factorized result for the decay amplitudes in rare radiative
$B$ decays at leading order in SCET, and at next-to-leading order in
$\alpha_s$.  
\end{abstract}

\maketitle

\section{Introduction}
The formalism of soft-collinear effective theory (SCET)
\cite{bauer1,bauer2,bauer3,bauer4,beneke1,beneke2} can be applied 
to high-energy processes in which massless quarks are emitted with
large energy. In this case a light quark with large energy has three
different scales $Q$, $Q\lambda$ and $Q\lambda^2$, where $Q$ is the
large energy scale and $\lambda \sim p_{\perp}/Q$ is a small
parameter. Here $p_{\perp}$ is a typical transverse momentum
scale. SCET has been applied to many high-energy processes in which
final-state particles include energetic light particles. For example,
the hard contribution to the  
heavy-to-light form factor \cite{bauer1,bauer2,beneke1,chay1,bs}, the
pion form factor and high-energy scattering \cite{ira}, and radative
leptonic $B$ decays $B \rightarrow \gamma e\nu$
\cite{lunghi,neubert1} have been studied. Recently it has been 
applied to nonleptonic $B$ decays into two light mesons by the authors 
\cite{chay2,chay3} and nonleptonic decays are shown to factorize at
leading order in SCET and to all orders in $\alpha_s$. 

In this paper we apply SCET to rare radiative $B$ decays $B\rightarrow
V\gamma$, where $V$ can be a $\rho$ meson or a $K^*$ meson.
In this case, employing a two-step
matching offers a convenient procedure to obtain the matrix elements
of the relevant operators evaluated below the scale $\mu_0 \sim
\sqrt{\Lambda_{\mathrm{QCD}} m_b}$. First we consider an effective
theory called $\mathrm{SCET}_{\mathrm{I}}$, in which the degrees of
freedom with $p\sim m_b$ are integrated out and the effective theory
describes physics below the scale $m_b$ and above $\mu_0$. In
$\mathrm{SCET}_{\mathrm{I}}$, we can construct all the effective
operators which are gauge invariant under collinear and ultrasoft
(usoft) gauge transformations. But a typical scale in
$\mathrm{SCET}_{\mathrm{I}}$ is of order $m_b \lambda \sim \sqrt{m_b
  \Lambda_{\mathrm{QCD}}}$, 
which is still too big for the evaluation of the matrix elements of
the effective operators. A next step is to go down to the second
effective theory $\mathrm{SCET}_{\mathrm{II}}$, in which all the
degrees of freedom with $p\sim \sqrt{m_b \Lambda_{\mathrm{QCD}}}$ are
integrated out and a typical scale in $\mathrm{SCET}_{\mathrm{II}}$ is
of order $\Lambda_{\mathrm{QCD}}$ and the small parameter becomes of
order $\Lambda_{\mathrm{QCD}}/m_b$. In $\mathrm{SCET}_{\mathrm{II}}$,
we can evaluate the matrix elements of the operators between meson
states. The two-step matching looks complicated, but the construction
and the power counting of operators become manifest.

Radiative $B$ decays have been previously considered in different 
theoretical frameworks. Ali and Parkhomenko \cite{ali} calculated hard
spectator contributions to radiative $B$ decays in the scheme of the
large energy effective theory, and presented their results at
next-to-leading order and to leading order in
$\Lambda_{\mathrm{QCD}}/m_b$. In this approach they calculated the
decay amplitudes in the full theory, and expanded them in powers of
$1/E$, where $E\sim m_b$. Radiative $B$ decays were also considered
in the context of the heavy quark mass limit \cite{seidel,bosch}, in
which the amplitudes in the full theory are expanded in powers of
$1/m_b$ \cite{bbns}, employing the light-cone
wave functions of the mesons convoluted with the hard scattering
amplitudes. The results in Refs.~\cite{ali,seidel,bosch} show that the
radiative $B$ decay amplitudes are factorized at next-to-leading order
in $\alpha_s$ and to leading order in $\Lambda_{\mathrm{QCD}}/m_b$. 
Subleading effects in radiative $B$ decays such as
the study of the topology of the weak annihilation and the
$W$-exchanges channels \cite{grinstein}, and the isospin breaking
effects \cite{kagan} were studied.

If we apply SCET to radiative $B$ decays, we can extend the proof of
the factorization theorem to all orders in $\alpha_s$ at leading order
in $\Lambda_{\mathrm{QCD}}/m_b$. It is possible because we require
that the effective operators in $\mathrm{SCET}_{\mathrm{II}}$ be gauge
invariant under collinear and soft gauge
transformations. The gauge invariance and the reparameterization
invariance put a serious constraint on the possible forms of the
effective operators in SCET \cite{scetop,bauer7}.
The gauge invariance is achieved by introducing the corresponding
Wilson lines, and these Wilson lines include the effects of collinear
and soft gluons to all orders in $\alpha_s$. The Wilson coefficients
of the effective operators can be calculated to  desired accuracy,
say, next-to-leading order, or next-to-next-to-leading order, but the
form of the effective operators remains the same.

Another advantage in applying SCET is that we can make the power
counting explicit at each step of the effective theories
$\mathrm{SCET}_{\mathrm{I}}$ and $\mathrm{SCET}_{\mathrm{II}}$. In
$\mathrm{SCET}_{\mathrm{I}}$, we can obtain gauge invariant operators
in powers of $\lambda$. When we integrate out an intermediate scale,
we need not worry about operator mixing in different powers of
$\lambda$ since the matching is perturbative. As we go down to
$\mathrm{SCET}_{\mathrm{II}}$, the small parameter now becomes $\sim
\Lambda_{\mathrm{QCD}}/m_b$ and the power counting is explicit. 
To summarize, our procedure to employ SCET is first to match QCD onto
$\mathrm{SCET}_{\mathrm{I}}$ at $\mu \sim m_b$, and factorize the
usoft-collinear interactions with the field redefintions. Finally we
match $\mathrm{SCET}_{\mathrm{I}}$ onto $\mathrm{SCET}_{\mathrm{II}}$
at $\mu_0 \sim \sqrt{m_b  \Lambda_{\mathrm{QCD}}}$, and evaluate the
matrix elements.

The organization of the paper is as follows: In Section II, we briefly
review the basic ingredients of SCET in studying radiative $B$
decays. We explain which types of operators are needed in computing
the decay amplitudes of radiative $B$ decays,
and the outline of the procedure in using SCET is sketched. In Section
III, all the relevant operators are derived in
$\mathrm{SCET}_{\mathrm{I}}$. There
are effective operators without spectator quarks, and if we include
the effects of spectator quarks, new four-quark operators
and four-quark operators with a photon are induced. These operators
are constructed in a gauge invariant way. In Section IV, the operators
obtained in $\mathrm{SCET}_{\mathrm{I}}$ are evolved down to
$\mathrm{SCET}_{\mathrm{II}}$, and we evaluate all the matrix elements
of the operators at leading order in SCET. In Section V, all the
contributions are combined to produce the decay amplitudes for the
radiative $B$ decays. In the final section, we
summarize all the effects of various operators and a conclusion is
presented. 

\section{Basics of SCET for $B\rightarrow V\gamma$ decays}
In order to describe radiative $B$ decays in SCET, the first step
is to construct effective operators in $\mathrm{SCET}_{\mathrm{I}}$ by
integrating out the degrees of freedom of order $m_b$ from the full
theory. The effective operators should be gauge invariant under
collinear and usoft gauge transformations, and these operators can be
systematically expanded in a power series of the small parameter
$\lambda$, of order $\sqrt{\Lambda_{\mathrm{QCD}}/m_b}$. 
After we construct all the operators contributing to radiative $B$
decays in $\mathrm{SCET}_{\mathrm{I}}$, we evolve these operators down
to $\mathrm{SCET}_{\mathrm{II}}$. In obtaining
$\mathrm{SCET}_{\mathrm{II}}$, we integrate all the off-shell modes of
order $\sqrt{m_b \Lambda_{\mathrm{QCD}}}$, and all the operators in
$\mathrm{SCET}_{\mathrm{II}}$ can be expanded in a power seris of the
small parameter, which  now 
becomes of order $\Lambda_{\mathrm{QCD}}/m_b$. At each step of the
effective theories, we match theories at the boundary requiring that
matrix elements of a given operator be the same in both theories. And
we go down to the scale of interest by using the renormalization group
equation. 

We begin with the effective Hamiltonian for $B$ decays in the full
QCD, which is given by
\begin{equation}
H_{\mathrm{eff}} = \frac{G_F}{\sqrt{2}} \sum_{p=u,c} V^*_{pd} V_{pb}
\Bigl( C_1 O_1^p +C_2 O_2^p +\sum_{i=3,\cdots,8} C_i O_i \Bigr),
\label{heff}
\end{equation}
where 
\begin{eqnarray}
O_1^p&=& (\overline{p}_i b_i)_{V-A} (\overline{d}_j p_j)_{V-A}, \
O_2^p = (\overline{p}_j b_i)_{V-A} (\overline{d}_i p_j)_{V-A},
\nonumber \\ 
O_3 &=& (\overline{d}_i b_i)_{V-A} \sum_q (\overline{q}_j q_j)_{V-A},
\ O_4 = (\overline{d}_j b_i)_{V-A} \sum_q (\overline{q}_i q_j)_{V-A},
\nonumber \\ 
O_5 &=& (\overline{d}_i b_i)_{V-A} \sum_q (\overline{q}_j q_j)_{V+A},
\ O_6 = (\overline{d}_j b_i)_{V-A} \sum_q (\overline{q}_i q_j)_{V+A},
\nonumber \\ 
O_7 &=& -\frac{em_b}{8\pi^2} \overline{d} \sigma_{\mu\nu} F^{\mu\nu}
(1+\gamma_5) b, \ O_8 = -\frac{gm_b}{8\pi^2} \overline{d}_i
\sigma_{\mu \nu} G^{\mu\nu}_a (T_a)_{ij} (1+\gamma_5)b_j.
\end{eqnarray}
Here $p$ is an up-type quark $u$ or $c$ quark, and $d$ is a down-type
quark $d$ (for $\overline{B}\rightarrow \rho \gamma$) or $s$ (for
$\overline{B} \rightarrow
K^* \gamma$) quark. The indices $i$, $j$ are color indices, 
$F^{\mu\nu}$ and $G^{\mu\nu}_a$ are the electomagnetic and the
chromomagnetic field strength tensors respectively, and $V_{ij}$ are
the Cabibbo-Kobayashi-Maskawa matrix elements. The sign
convention for $O_{7,8}$ corresponds to negative $C_{7,8}$ and the
covariant derivative is defined as
$D^{\mu} = \partial^{\mu} -igT_a A^{a\mu} -ieQ \mathcal{A}^{\mu}$,
where $A^{a\mu}$ ($\mathcal{A}^{\mu}$) is the gluon (photon) field.

The leading contribution to $B\rightarrow V\gamma$ comes from the
operator $O_7$ in the full theory. In the effective theory, the
leading-contribution comes from the leading-order operator derived
from $O_7$. In the following computation, we
will neglect the radiative corrections of the 
penguin operators since their coefficients are proportional to
$\alpha_s C_i$ ($i=3,\cdots , 6$), which are numerically small
compared to $C_7$. But we include the loop correction from $O_1$ since
$\alpha_s C_1$ is not negligible compared to $C_7$. When the penguin
operators appear at leading order in $\alpha_s$, we include their
effects. 

Before we consider the effective operators for radiative $B$ decays in
$\mathrm{SCET}_{\mathrm{I}}$,  let us introduce the notations we use
in this paper and discuss the power counting of operators. We choose
the direction of the vector meson as $n^{\mu}$, and the direction of
the photon as $\overline{n}^{\mu}$. And the vector mesons are regarded
as massless at leading order in SCET. Therefore the $SU(3)$ breaking
effects in $\overline{B} \rightarrow \rho \gamma$ and $\overline{B}
\rightarrow K^* \gamma$ and the $SU(2)$ isospin
breaking effects in $\overline{B}\rightarrow \rho^0 \gamma$
and $B^- \rightarrow \rho^- \gamma$  arise at subleading order. The
quark mass effect in SCET was discussed in Ref.~\cite{wise}, which can
cause leading operators in $\mathrm{SCET}_{\mathrm{II}}$. However the
matrix elements of these operators in radiative $B$ decays at leading
order in SCET turn out to vanish. This is discussed in Appendix in
detail.

We denote the collinear fields $\xi$ and $\chi$ as the collinear
fields in the $n^{\mu}$ and the $\overline{n}^{\mu}$ directions
respectively, satisfying  
\begin{equation}
\FMslash{n} \xi =0, \ \frac{\FMslash{n} \overline{\FMslash{n}}}{4} \xi
=\xi, \ \overline{\FMslash{n}} \chi =0, \
\frac{\overline{\FMslash{n}}\FMslash{n}}{4} \chi = \chi. 
\end{equation}
The collinear gluon field $A_n^{\mu}$ in the $n^{\mu}$ direction can
be decomposed as 
\begin{equation}
A_n^{\mu} = \frac{n^{\mu}}{2} \overline{n}\cdot A_n + A_{n\perp}^{\mu}
+\frac{\overline{n}^{\mu}}{2} n\cdot A_n = \mathcal{O} (\lambda^0)
+\mathcal{O} (\lambda) + \mathcal{O} (\lambda^2),
\label{cgluon}
\end{equation}
and the power counting of the operators involving
collinear fields can be made explicit as shown in
Ref.~\cite{bauer2,bauer4}. On the other hand, the photon field 
$\mathcal{A}^{\mu}$ cannot be decomposed as the collinear gluon to
facilitate the power counting. The decomposition of the collinear
gluon field in Eq.~(\ref{cgluon}) is meaningful since quarks and
gluons have small fluctuations due to strong interaction. If we
consider an external photon, the photon field does not have such small
fluctuations and the momentum of the photon is always of order
$\lambda^0$, and only the transverse part contributes to decay
amplitudes. Therefore we do not include the photon field in the power
counting of operators. Though we assign any power of $\lambda$ to
the photon field, it does not affect the relative power counting of
various operators since all the operators for $B\rightarrow
V\gamma$ will include a single photon.

There are two kinds of effective operators in
$\mathrm{SCET}_{\mathrm{I}}$ from different channels, which
can be classified into the operators without spectator quarks and
those with spectator quarks. 
The leading operator for the first type of operators comes from $O_7$
by replacing the quarks in the full theory by those in the
effective theory and making the resultant operator gauge invariant
under collinear gauge transformations. We also have the contributions
from the operators $O_1$ and $O_8$ by radiative corrections, whose
matrix elements are proportional to the matrix element of $O_7$. The
resultant operators are heavy-light quark bilinear operators with a
photon field. These operators, after we make them gauge invariant by
attaching Wilson lines, can produce operators with an additional
external gluon. At order $g$, these operators can be obtained by
attaching a collinear gluon to $O_7$, or a photon to $O_8$, or a gluon
and a photon to $O_1$.

All these operators without spectator quarks can be obtained by
attaching a photon or a gluon to fermion lines in the operators $O_7$,
$O_1$ and $O_8$. When we attach a photon or a gluon to each
operator, if any intermediate states have momenta of
order $\lambda^0$, we integrate out these modes to obtain effective
operators. If the intermediate states have momenta of order $\lambda
m_b$, then these modes cannot be integrated out in
$\mathrm{SCET}_{\mathrm{I}}$. Instead that channel should be considered
in $\mathrm{SCET}_{\mathrm{II}}$. In order to make
the operators gauge invariant under collinear gauge
transformations, we also have to consider attaching collinear gluons
and integrate out all the off-shell modes of order $\lambda^0$. We
compute all the operators in $\mathrm{SCET}_{\mathrm{I}}$ to order
$\lambda$ in order to
obtain the leading-order result due to the possible enhancement in
evolving the operators to $\mathrm{SCET}_{\mathrm{II}}$.

The second type of the operators is the operators with spectator
quarks. These include the four-quark operators of  
the generic form $\overline{\chi}\Gamma_1 h \cdot \overline{\xi}
\Gamma_2\xi$, and $\overline{\xi} \Gamma_1 h \cdot \overline{\chi}
\Gamma_2 \xi$. As mentioned in Ref.~\cite{chay3}, we can use
the Fierz transformation to make the operator $\overline{\xi} \Gamma_1
h \cdot \overline{\chi} \Gamma_2 \xi$ of the form
$\overline{\chi}\Gamma_1 h \cdot \overline{\xi} \Gamma_2\xi$, and we
will consider the four-quark operators only in the form
$\overline{\chi}\Gamma_1 h \cdot \overline{\xi} 
\Gamma_2\xi$ from now on. In these four-quark operators, the spectator
quark is a collinear quark $\chi$. And we can make  time-ordered
products of these operators with the electromagnetic interaction with
a collinear quark $\chi$ and an usoft quark. It is necessary to
include the four-quark operators of the form $\overline{\chi}\Gamma_1
h \cdot \overline{\xi}  \Gamma_2\xi$ in $\mathrm{SCET}_{\mathrm{I}}$
since the time-ordered products of the four-quark operators with the
electromagnetic interaction contributes to radiative $B$ decays.

There are also four-quark operators with a photon field
of the form  $\overline{q}_{us} \Gamma_1 h \cdot \overline{\xi}
\Gamma_2 \xi \cdot \mathcal{A}^{\mu}$, where $\Gamma_i$ are Dirac
matrices. These operators are obtained from the four-quark operators
in Eq.~(\ref{heff}) by attaching a photon to all the quarks except the
spectator quark of the $B$ meson. In this case, the intermediate state
has momentum of $m_b$ and can be integrated out to produce effective
operators. The effects of the operators with a
photon attached to a spectator quark are already included in the
four-quark operators mentioned above. In this case
the intermediate state has the momentum of order $m_b \lambda$ and
cannot be integrated out in $\mathrm{SCET}_{\mathrm{I}}$.

When we evaluate the matrix elements of all the operators in
$\mathrm{SCET}_{\mathrm{I}}$, we consider 
the matrix elements of the operators without spectator quarks, the
four-quark operators with a photon, and the time-ordered products of
the four-quark operators with
the electromagnetic interaction of an usoft quark and a collinear
quark $\chi$. In addition, when we consider the hard scattering
contribution to the form factor for $B\rightarrow V$, we also include
the time-ordered products of the operators without spectator quarks
with the subleading collinear effective Lagrangian. And these matrix
elements are evolved down to $\mathrm{SCET}_{\mathrm{II}}$. By
decoupling soft gluon 
interactions, we obtain gauge invariant sets of operators and
time-ordered products under collinear and soft gauge
transformations, and evaluate the matrix elements. 
  
We can consider other forms of four-quark operators, say, 
$\overline{\xi}\Gamma_1 h \cdot \overline{\xi} \Gamma_2\xi
\mathcal{A}^{\mu}$, in which all the light quarks are collinear fields
in the $n^{\mu}$ direction. We can make time-ordered products of these
operators with the interaction of collinear gluons with an usoft quark
and a collinear quark $\xi$ \cite{hill,bauer7}. If
there are no external gluons in the time-ordered products, all the
collinear gluons should appear as internal gluons, and the
final operators correspond to, at higher orders in
$\alpha_s$, the four-quark operators with a photon of the form
$\overline{q}_{us} \Gamma_1 h \cdot \overline{\xi} \Gamma_2 \xi \cdot
\mathcal{A}^{\mu}$ after we perform the time-ordered product with the
interaction involving the usoft quark. If there
are external gluons, these are collinear gluons in the $n^{\mu}$
direction. In this case, these gluons should belong to
the final state of a vector meson. Therefore
it involves higher Fock space for the vector meson state, which we
will neglect since the contributions are subleading.  

\section{Operators in $\mbox{\boldmath $\mathrm{SCET}_{\mathrm{I}}$
    \unboldmath}$} 
\subsection{Operators without spectator quarks}
\begin{figure}[b]
\begin{center}
\epsfig{file=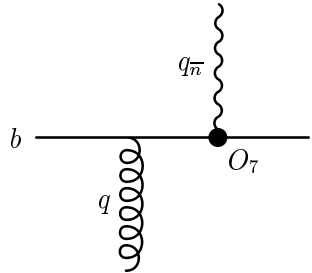, width=4.0cm}
\end{center}
\vspace{-0.5cm}
\caption{QCD diagram attaching a collinear gluon to the $b$ quark to
  make a gauge invariant operator. A wavy line represents a photon,
  and a curly line represents a gluon. The momentum of the photon
  (gluon) is outgoing (incoming).}  
\label{fig1}
\end{figure}

In radiative $B$ decays, the leading operator is
$O_7$ and the corresponding gauge invariant operators in
$\mathrm{SCET}_{\mathrm{I}}$ to first order in $\lambda$ are given by
\begin{equation}
O_7 \rightarrow A_7^{(0)} O_7^{(0)} + A_7^{(1a)} O_7^{(1a)} +
A_7^{(1b)} O_7^{(1b)},
\label{o7op}
\end{equation}
where $A_7^{(0,1a,1b)}$ are the Wilson coefficients, which are
operators in SCET. The operator
$O_7^{(0)}$ is the leading operator, and $O_7^{(1a,1b)}$ are the
subleading operators suppressed by $\lambda$ compared to
$O_7^{(0)}$. These operators are obtained by attaching a collinear gluon
$A_n^{\mu}$ to the heavy quark and integrate out off-shell modes, as
shown in Fig.~\ref{fig1}. The operators are given as
\begin{eqnarray}
O_7^{(0)} &=&  \frac{em_b^2}{8\pi^2}  \Bigl(\overline{\xi} W \Bigr)
\FMslash{\overline{n}} \FMSlash{\mathcal{A}} (1+\gamma_5) h =
-\frac{em_b^2}{4\pi^2} \Bigl( \overline{\xi} W \Bigr)
\FMSlash{\mathcal{A}} (1-\gamma_5) h, \nonumber \\
O_7^{(1a)} &=& \frac{em_b}{4\pi^2} \Bigl(\overline{\xi} W \Bigr)
\FMSlash{\mathcal{A}} [W^{\dagger} i\FMSlash{D}_{n\perp} W]
(1+\gamma_5)h, \nonumber \\
O_7^{(1b)} &=& \frac{em_b}{4\pi^2} \Bigl(\overline{\xi} W\Bigr)
[W^{\dagger}   i\FMSlash{D}_{n\perp} W]
\FMSlash{\mathcal{A}}(1+\gamma_5)  h, 
\label{o7op2}
\end{eqnarray}
where $\mathcal{A}^{\mu}$ is the photon field, and $h$ is the heavy
quark field in the heavy quark effective theory. The factor $W$ is the
Wilson line defined as  
\begin{equation}
W=\sum_{\mathrm{perm}} \exp \Bigl[-g \frac{1}{\overline{n} \cdot
    \mathcal{P}} \overline{n} \cdot A_n \Bigr],
\end{equation}
where $\mathcal{P}^{\mu} = \overline{n}\cdot \mathcal{P} n^{\mu}/2
+\mathcal{P}_{\perp}^{\mu}$ is the label momentum operator for
collinear fields in the $n^{\mu}$ direction.

When we actually compute the Feynman diagram in Fig.~\ref{fig1}, we
obtain $O_7^{(0)}$ and $O_7^{(1a)}$ at tree level. However,
the operators $O_7^{(1a,1b)}$ are the most
general operators at subleading order considering the Dirac structure
and the collinear gauge invariance. When we match SCET to the
full theory at tree level, the Wilson coefficient for $O_7^{(1b)}$ is
zero, but in the matching at higher order in $\alpha_s$, a nonzero
coefficient $A_7^{(1b)}$ can be developed. If we calculate decay
amplitudes to next-to-leading order accuracy, we need to calculate
$A_7^{(0)}$ to order $\alpha_s$ and $A_7^{(1a,1b)}$ to order
$\alpha_s^0$ since the operators $O_7^{(1a,1b)}$ begin with order $g$
and these operators will be combined with the effective Lagrangian in
the time-ordered products, which are of order $\alpha_s$.  
There are no other subleading operators by including the subleading
correction to the collinear field $\xi$ in $O_7^{(0)}$, where the
fermion field $\psi$ in the full theory changes to
\begin{equation}
\psi(x) \rightarrow \Bigl( 1+\frac{1}{\overline{n} \cdot (\mathcal{P}
 +gA_n)} (\FMSlash{\mathcal{P}}_{\perp}+g\FMSlash{A}_{n\perp})
 \frac{\FMslash{\overline{n}}}{2} \Bigr) \xi.
\end{equation}
 It is because the subleading 
correction to the collinear field $\xi$ is proportional to
$\FMslash{\overline{n}}$ and this vanishes when contracted with
$\FMslash{\overline{n}}$ in the operator $O_7^{(0)}$.

The Wilson coefficient
$A_7^{(0)}$, to first order in $\alpha_s$, is given as  
\begin{eqnarray}
A_7^{(0)} (\mu) &=&  1-\frac{\alpha_s C_F}{4\pi} \Bigl[ \frac{1}{2}
  \ln^2 \frac{\mu^2}{m_b^2} +\frac{7}{2} \ln \frac{\mu^2}{m_b^2} -2
  \ln \frac{\mu^2}{m_b^2} \ln \overline{n} \cdot
  \hat{\mathcal{P}^{\dagger}}   \nonumber \\
&&+2\ln^2
  \overline{n} \cdot \hat{\mathcal{P}^{\dagger}} +2 \mathrm{Li}_2
  (1-\overline{n}  \cdot \hat{\mathcal{P}^{\dagger}}) -2\ln
  \overline{n} \cdot   \hat{\mathcal{P}^{\dagger}} 
  +\frac{\pi^2}{12} +6 \Bigr],
\label{a70}
\end{eqnarray}
where $\overline{n} \cdot \hat{\mathcal{P}} = \overline{n} \cdot
\mathcal{P}/m_b$, and $\mathrm{Li}_2 (x)$ is the dilogarithmic
function. The coefficient $A_7^{(0)}$ is the same as $C_9$ in
Ref.~\cite{bauer3}. The remaining Wilson coefficients to zeroth
order in $\alpha_s$ are given by
\begin{equation}
A_7^{(1a)}=1, \ \ A_7^{(1b)}=0.
\end{equation}
Since the Wilson coefficients are operators, $A_7^{(0)}$ in
Eq.~(\ref{o7op}) means that the Wilson coefficient is sandwiched by
the operator, that is, it implies that
\begin{equation} 
A_7^{(0)} O_7^{(0)} = \frac{em_b^2}{8\pi^2} \overline{\xi} W
\FMslash{\overline{n}} A_7^{(0)} (\overline{n} \cdot
\mathcal{P}^{\dagger}) 
\FMSlash{\mathcal{A}} (1+\gamma_5) h=A_7^{(0)} (\overline{n} \cdot p,
\overline{n} \cdot q)
\frac{em_b^2}{8\pi^2} \overline{\xi} W 
\FMslash{\overline{n}}\FMSlash{\mathcal{A}} (1+\gamma_5) h,
\end{equation}
where $p$ ($q$) is the momentum of the quark $\xi$ (the gluon in
$W$).

We can obtain operators from $O_1$ and
$O_8$, proportional to $O_7$ by attaching an external photon and a
gluon in a loop, whose matrix elements are proportional to the matrix
element of  $O_7$ through the radiative corrections. The leading-order
contributions from the operators $O_1$ and $O_8$, which are
proportional to $O_7^{(0)}$ will be included in calculating the decay
amplitudes in Section IV. 

\begin{figure}[b]
\begin{center}
\epsfig{file=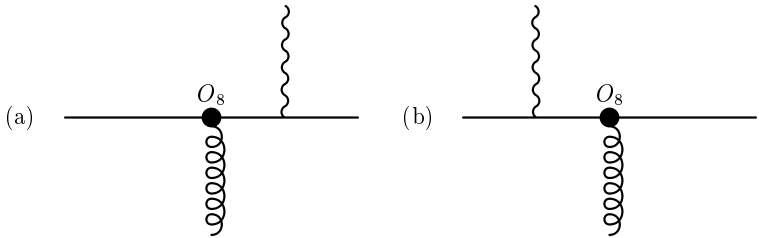, width=10.0cm}
\end{center}
\vspace{-0.5cm}
\caption{QCD diagrams attaching a photon to each fermion to the
operator $O_8$ in order to generate the operator with an external
photon and an external gluon. The momentum  of the photon (gluon) is
outgoing (incoming). Diagram (a) is of order $\lambda$ compared to
$O_7^{(0)}$. Diagram (b) is suppressed by $\lambda^2$ and is
neglected.}   
\label{fig2}
\end{figure}

Now let us consider the operators which consist of a heavy-to-light
current with an external photon and an external gluon, in addition to
$O_7^{(1a,1b)}$. From the operator $O_8$ in the full theory, we can
construct such an operator. It is given by 
\begin{equation}
O_8 \rightarrow Q_d A_8^{(1a)} O_7^{(1a)} + Q_d A_8^{(1b)} O_7^{(1b)},
\end{equation}
where $Q_d$ is the electric charge of the down-type quark ($d$ or
$s$). Note that there are only two independent operators
$O_7^{(1a,1b)}$ at order $\lambda$ in
$\mathrm{SCET}_{\mathrm{I}}$. Here the operator without a photon from
$O_8$ is not included since it does not contribute to radiative $B$
decays, but we comment that the leading operator from $O_8$ without a
photon field is suppressed 
by $\lambda^2$ compared to $O_7^{(0)}$. We can compute the Wilson
coefficients $A_8^{(1a,1b)}$ to any desired accuracy, but at
next-to-leading order, we need the tree-level 
Wilson coefficients since the operators begin with order $g$. Matching
the operators with the full theory at tree level, as shown in
Fig.~\ref{fig2}, the Wilson coefficient $A_8^{(1b)}$ is zero, and
$Q_d A_8^{(1a)} O_7^{(1a)}$ can be written as
\begin{equation}
Q_d A_8^{(1a)} O_7^{(1a)}= \frac{eQ_d m_b}{4\pi^2} \overline{\xi} W
\FMSlash{\mathcal{A}} \frac{1}{\overline{n} \cdot
  \mathcal{P}^{\dagger}} \overline{n} \cdot \mathcal{P}
  [W^{\dagger}i\FMSlash{D}_{n\perp} W] 
  (1+\gamma_5) h,
\label{o8gp}
\end{equation}
where $Q_d$ is the electric charge of the collinear quark $\xi$. 
The operator in Eq.~(\ref{o8gp}) is obtained by attaching a photon to
fermion lines in the operator $O_8$, as shown in Fig.~\ref{fig2}. Note
that the contribution of Fig.~\ref{fig2} (b) is suppressed by $\lambda$
compared to that of Fig.~\ref{fig2} (a), and it is neglected here. 

\begin{figure}[b]
\begin{center}
\epsfig{file=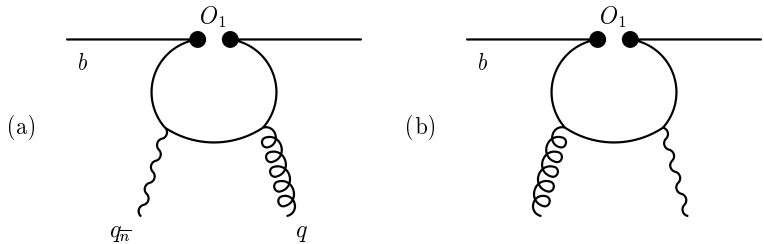, width=10.0cm}
\end{center}
\vspace{-0.5cm}
\caption{QCD diagrams generating the operator with an external photon
  and an external gluon from $O_1$, in which a
  photon and a gluon are attached to the internal fermion
  line. The gluon momentum $q$ is incoming and the photon momentum
  $q_{\overline{n}}$ is outgoing. The same diagrams with the penguin
  operators are neglected.} 
\label{fig3}
\end{figure}

We can also obtain an operator with a photon and a gluon from the
operator $O_1$. We cannot make the operator from $O_2$ due to the
color structure, and we neglect the contributions of penguin
operators.  The full QCD diagram is shown in Fig.~\ref{fig3} and
the resultant operator can be written as \cite{greub,hurth}
\begin{equation}
\overline{d} \mathcal{A}^{\mu} A^{\nu} I_{\mu\nu} b,
\label{o1a}
\end{equation}
where
\begin{eqnarray}
I_{\mu\nu} &=& \frac{geQ_u}{8\pi^2} \Bigl[i\epsilon_{\mu
    \nu\alpha\beta} \Bigl( q^{\alpha} \Delta i_5 +q_{\overline{n}}^{\alpha}
    \Delta i_6 \Bigr) \nonumber \\
&&+\frac{1}{q\cdot q_{\overline{n}}} \bigl( i\epsilon_{\rho\nu
    \alpha\beta} q_{\mu} q^{\rho} q_{\overline{n}}^{\alpha}  \Delta
    i_{23} +i \epsilon_{\rho\mu\alpha \beta} q^{\rho}
    q_{\overline{n}}^{\alpha} q_{\overline{n}\nu} \Delta i_{26} \Bigr)
    \Bigr] \gamma^{\beta} (1-\gamma_5).
\end{eqnarray}
Here the momentum of the gluon $q$ is incoming to the loop and the
momentum of the photon $q_{\overline{n}}$ is outgoing. 
The quantities $\Delta i_n \equiv \Delta i_n (z_0, z_1, z_2)$ are
functions of the three variables $z_0$, $z_1$ and $z_2$, which are 
the invariant mass-squared of the fermion-antifermion pair in the
loop, the gluon momentum squared, and the photon momentum squared,
each divided by the internal fermion mass $m_i^2$, and they are given
by 
\begin{equation}
z_0 = \frac{(q-q_{\overline{n}})^2}{m_i^2} = \frac{(p_1 - m_b
  v)^2}{m_i^2} =\frac{m_b^2}{m_i^2} (1-u),  \ z_1 = \frac{q^2}{m_i^2},
\ z_2 = \frac{q_{\overline{n}}^2}{m_i^2}=0,
\end{equation}
where $u$ is the momentum fraction of the $d$ quark in a vector
meson. Replacing the quark fields with the fields in SCET, it turns
out that $I_{\mu\nu}$ starts from terms of order $\lambda$ apart from
$\Delta i_n$'s. Therefore we need to expand $\Delta i_n$ to leading
order in $\lambda$, which are given by $\Delta i_n (z_0,0,0)$. In this
case, $\Delta i_n$ have the relation
\begin{equation}
\Delta i_{23} (z_0,0,0) = \Delta i_{26}(z_0,0,0) = \Delta
i_6(z_0,0,0) = \Delta i_5(z_0,0,0),
\label{dis}
\end{equation}
and $\Delta i_5 (z_0,0,0)$ is given by
\begin{equation}
\Delta i_5 (z_0,0,0) = -\frac{4}{z_0} \Bigl[ \mathrm{Li}_2 \Bigl(
  \frac{2}{1-\sqrt{1- 4/z_0 + i\epsilon}} \Bigr)  + \mathrm{Li}_2 \Bigl(
  \frac{2}{1+\sqrt{1- 4/z_0 + i\epsilon}} \Bigr) \Bigr]+2.
\label{ifive}
\end{equation}
Note that, in Ref.~\cite{hurth}, the momenta of the gluon and the
photon are both outgoing, while the momentum of the gluon is reversed
in our case in order to keep a consistent convention that a collinear
gluon is incoming. Because of this different choice of the directions
of the momenta, $\Delta i_5$ and $\Delta i_{23}$ in Eqs.~(\ref{dis})
and (\ref{ifive}) have opposite signs compared to those in
Ref.~\cite{hurth}.  

Replacing the fields in Eq.~(\ref{o1a}) by the fields in the effective
theory, and making the resultant operator collinear gauge invariant,
we obtain the effective operators at order $\lambda$, which are 
given by
\begin{equation}
O_1 \rightarrow Q_u A_1^{(1a)} O_7^{(1a)} +Q_u A_1^{(1b)} O_7^{(1b)},
\label{o1op}
\end{equation}
where $Q_u$ is the electric charge of the up-type quark in the fermion
loop. To next-to-leading order, the Wilson
coefficients are given by
\begin{equation} 
A_1^{(1a)} = -A_1^{(1b)} = \frac{1}{4} H(\overline{n} \cdot
\mathcal{P}^{\dagger}, s_i),
\end{equation}
where $H (\overline{n} \cdot \mathcal{P}^{\dagger}, s_i) =\Delta i_5
(z_0 \rightarrow (1-\overline{n} \cdot \mathcal{P}^{\dagger}/m_b)/s_i,
0,0)$ is now an operator with $s_i = m_i^2/m_b^2$. Compared to
$h(u,s)$ in Ref.~\cite{bosch}, $H(u,s) = -u h(u,s)$. The operators in
Eq.~(\ref{o1op}) can be written as
\begin{eqnarray}
 Q_u A_1^{(1a)} O_7^{(1a)} +Q_u A_1^{(1b)} O_7^{(1b)}&=&\frac{eQ_u
   m_b}{16\pi^2}  
\Bigl(\overline{\xi} W H
  (\overline{n} \cdot \mathcal{P}^{\dagger}, s_i) 
  \FMSlash{\mathcal{A}} [W^{\dagger} i\FMSlash{D}_{n\perp} W]
  (1+\gamma_5) h \nonumber \\
&&-\overline{\xi} W H (\overline{n} \cdot \mathcal{P}^{\dagger}, s_i)
     [W^{\dagger}   i\FMSlash{D}_{n\perp} W] 
  \FMSlash{\mathcal{A}} (1+\gamma_5) h \Bigr).
\label{o1gp}
\end{eqnarray}

\begin{figure}[b]
\begin{center}
\epsfig{file=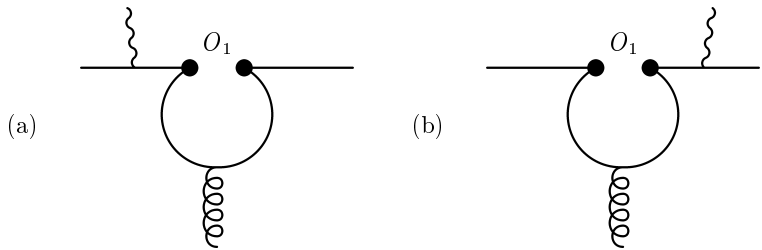, width=10.0cm}
\end{center}
\vspace{-0.5cm}
\caption{QCD diagrams generating an operator with a gluon and a photon
  from $O_1$. The resultant operator is suppressed by $\lambda^2$
  compared to $O_7^{(1a)}$, and is neglected.}
\label{fig4}
\end{figure}

So far, we have considered the effective operators from the operator
$O_1$, in which a photon and a gluon are attached to the internal
fermion loop. However, there
are other possibilities to attach a photon and a gluon to
fermions. For example, we can attach a gluon in the fermion loop and a
photon to external fermions, as shown in Fig.~\ref{fig4}. 
Without the photon, the fermion loop with a gluon attached to it
produces an operator \cite{ali2} 
\begin{equation}
\frac{g}{16\pi^2} \Bigl(\frac{4}{3} \ln \frac{m_i}{\mu} +\frac{2}{3}
+4B \Bigr)  \overline{d}  (q^2
\gamma_{\mu} -q_{\mu} \FMslash{q} ) A^{\mu} (1-\gamma_5) b,
\label{o1glu}
\end{equation}
with
\begin{equation}
B= \int_0^1 dz \,  z (1-z) \ln \Bigl( 1 -z (1-z) \frac{q^2}{m_i^2} \Bigr),
\end{equation}
where $m_i$ is the fermion mass in the loop and $q^{\mu}$ is the gluon
momentum. For the on-shell collinear gluon $q^2=0$, when we replace
the quark fields by the fields in the effective theory, the operator
in Eq.~(\ref{o1glu}) is proportional to
$\overline{\xi} q\cdot A_n \FMslash{q}_{\perp}
 (1-\gamma_5)h$, and it is suppressed by
$\lambda^3$ compared  to $O_7^{(0)}$ since $q\cdot A_n
q_{\perp}^{\mu}\sim O(\lambda^3)$. When we attach a photon to
external fermions to make the final effective operators, the power
counting does not change, hence they are neglected. 

\begin{figure}[t]
\begin{center}
\epsfig{file=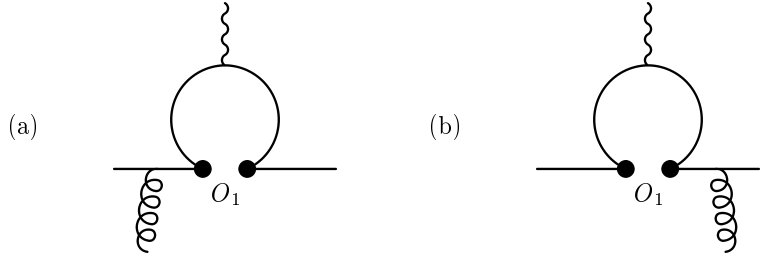, width=10.0cm}
\end{center}
\vspace{-0.5cm}
\caption{QCD diagrams in which a photon attached to the internal
  fermion line and a gluon is attached to one of the  external
  fermions of the operator   $O_1$. These diagrams 
  vanish due to the electromagnetic gauge invariance.}
\label{fig5}
\end{figure}
The remaining possibility is to attach a photon to the fermion loop
and a gluon is attached to external fermions of the operator $O_1$, as
shown in Fig.~\ref{fig5}. This operator is zero
due to the electromagnetic gauge invariance. That is, the fermion loop
in Fig.~\ref{fig5} is proportional to Eq.~(\ref{o1glu}), replacing the
gluon field by the photon field. Since the photon is on-shell, $q^2=0$
and $q\cdot \mathcal{A} =0$, the operator vanishes.

\subsection{Operators of the form $(\overline{\chi}\Gamma_1
  h)(\overline{\xi} \Gamma_2 \xi)$ }
The operators with spectator quarks involve
four-quark operators and four-quark operators with a photon. The
four-quark operators are generically of the form
$(\overline{\chi}\Gamma_1  h)(\overline{\xi} \Gamma_2 \xi)$.
The inclusion of these operators in $\mathrm{SCET}_{\mathrm{I}}$ is
necessary because there is 
an electromagnetic interaction of a photon in the $\overline{n}^{\mu}$
direction with an usoft quark $q_{us}$ and a collinear quark
$\chi$. The time-ordered products of the four-quark operators with the
electromagnetic interaction correspond to the emission of an energetic
photon from the spectator quark. At first sight, we can make an
operator by attaching a photon to a spectator quark and integrate out
the intermediate state to make an operator, but
the intermediate state has the momentum of order $m_b \lambda$, which
cannot be integrated out in $\mathrm{SCET}_{\mathrm{I}}$. Therefore we
construct the four-quark operators of the form $(\overline{\chi}\Gamma_1 
h)(\overline{\xi} \Gamma_2 \xi)$ in $\mathrm{SCET}_{\mathrm{I}}$ and
make the time-ordered products of these operators with the
electromagnetic interaction. Note
that, if a photon is attached to external fermions other than the
spectator quark, the intermediate state has the momentum of order
$m_b$, and it can be integrated out to produce four-quark operators
with a photon. This will be considered in the next subsection.

The explicit form of the effective four-quark operators can be
obtained from the four-quark operators for nonleptonic decays in
Ref.~\cite{chay3} with a minor modification. We 
only have to change $n^{\mu} \leftrightarrow \overline{n}^{\mu}$ and
$\chi \leftrightarrow \xi$ in the effective operators in
Ref.~\cite{chay3}. These operators are given as 
\begin{eqnarray}
O_{1R} &=& \Bigl( (\overline{\chi}^u \overline{W})_{\alpha} 
h_{\alpha} \Bigr)_{V-A} \Bigl( (\overline{\xi}^d W)_{\beta}
(W^{\dagger}\xi^u)_{\beta} \Bigr)_{V-A}, \nonumber \\
O_{2R} &=&  \Bigl( (\overline{\chi}^u
\overline{W} )_{\beta} h_{\alpha} \Bigr)_{V-A} \Bigl(
(\overline{\xi}^d W )_{\alpha} (W^{\dagger}\xi^u)_{\beta}
\Bigr)_{V-A}, \nonumber \\ 
O_{3R} &=& \Bigl( (\overline{\chi}^d \overline{W})_{\alpha} h_{\alpha}
\Bigr)_{V-A} \sum_q \Bigl( (\overline{\xi}^q
W)_{\beta} (W^{\dagger}\xi^q)_{\beta}
\Bigr)_{V-A}, \nonumber \\ 
O_{4R} &=&  \Bigl( (\overline{\chi}^d
\overline{W} )_{\beta} h_{\alpha} \Bigr)_{V-A} \sum_q \Bigl(
(\overline{\xi}^q W )_{\alpha} 
(W^{\dagger}\xi^q)_{\beta} \Bigr)_{V-A}, \nonumber \\
O_{5R} &=& \Bigl( (\overline{\chi}^d \overline{W})_{\alpha} h_{\alpha}
\Bigr)_{V-A} \sum_q \Bigl( (\overline{\xi}^q 
W)_{\beta} (W^{\dagger} \xi^q)_{\beta}
\Bigr)_{V+A}, \nonumber \\ 
O_{6R} &=&  \Bigl( (\overline{\chi}^d
\overline{W})_{\beta} h_{\alpha} \Bigr)_{V-A} \sum_q \Bigl(
(\overline{\xi}^q W )_{\alpha} 
(W^{\dagger}\xi^q)_{\beta} \Bigr)_{V+A}, \nonumber \\
O_{1C} &=&  \Bigl( (\overline{\chi}^d
\overline{W} )_{\beta} h_{\alpha} \Bigr)_{V-A} \Bigl(
(\overline{\xi}^u W )_{\alpha} 
(W^{\dagger}\xi^u)_{\beta} \Bigr)_{V-A}, \nonumber \\
O_{2C} &=& \Bigl( (\overline{\chi}^d \overline{W})_{\alpha} h_{\alpha}
\Bigr)_{V-A} \Bigl( (\overline{\xi}^u W)_{\beta} 
(W^{\dagger}\xi^u)_{\beta} \Bigr)_{V-A}, \nonumber \\
O_{3C} &=&  \sum_q \Bigl( (\overline{\chi}^q
\overline{W} )_{\beta} h_{\alpha} \Bigr)_{V-A}  \Bigl(
(\overline{\xi}^d W  )_{\alpha} 
(W^{\dagger}\xi^q)_{\beta} \Bigr)_{V-A}, \nonumber \\
O_{4C} &=& \sum_q \Bigl( (\overline{\chi}^q \overline{W})_{\alpha}
h_{\alpha} \Bigr)_{V-A}  \Bigl( (\overline{\xi}^d
W)_{\beta} (W^{\dagger} \xi^q)_{\beta} \Bigr)_{V-A}.
\label{fquark}
\end{eqnarray}
Here the summation over $q$ goes over to light massless quarks, say,
$u$, $d$, and $s$ quarks, and $\overline{W}$ is the Wilson line which
is given by
\begin{equation}
\overline{W}=\sum_{\mathrm{perm}} \exp \Bigl[-g \frac{1}{n \cdot
    \mathcal{Q}} n \cdot A_{\overline{n}} \Bigr],
\end{equation}
where $\mathcal{Q}^{\mu} = n\cdot \mathcal{Q} \overline{n}^{\mu}/2
+\mathcal{Q}_{\perp}^{\mu}$ is the label momentum operator for the 
collinear fields in the $\overline{n}^{\mu}$ direction. The operators
in Eq.~(\ref{fquark}) with $\overline{W}$ and $W$ are collinear gauge
invariant.  The Wilson coefficients of the operators in
Eq.~(\ref{fquark}) are the same as the corresponding operators in
Ref.~\cite{chay3}.  The time-ordered products of these operators
with the electromagnetic 
interaction with an usoft quark and a collinear quark $\chi$
contributes to the radiative $B$ decays. These correspond to the
contribution from the annihilation and the $W$-exchange channels. 

\begin{figure}[t]
\begin{center}
\epsfig{file=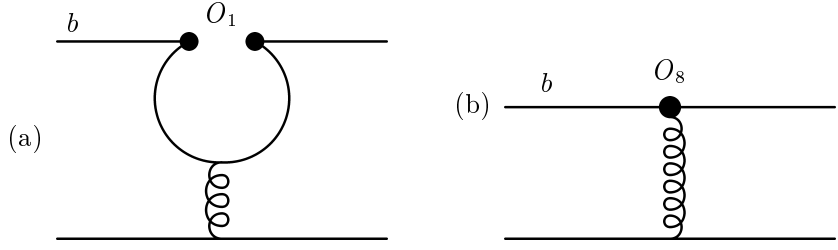, width=11.0cm}
\end{center}
\vspace{-0.5cm}
\caption{QCD diagrams generating the four-quark operators
  from $O_1$ and $O_8$. But these operators are included in the
  four-quark operators  in Eq.~(\ref{fquark}).}
\label{fig6}
\end{figure}

The operators $O_1$ and $O_8$ can induce four-quark operators, as
shown in Fig.~\ref{fig6}. However, these operators are 
four-quark operators which can
be expressed in terms of the four-quark operators in
Eq.~(\ref{fquark}). We include these operators in the operators given by
Eq.~(\ref{fquark}), and their effects appear in the
effective Wilson coefficients of the operators.

\subsection{Four-quark operators with a photon field}
\begin{figure}[b]
\begin{center}
\epsfig{file=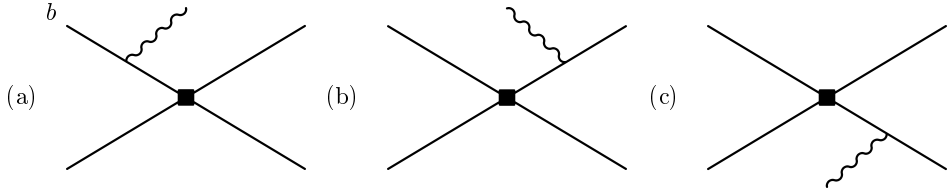, width=12.5cm}
\end{center}
\vspace{-0.5cm}
\caption{The four-quark operator with a
  photon in $\mathrm{SCET}_{\mathrm{I}}$ arising from the four-quark
  operators, which correspond to annihilation channels or $W$-exchange
  channels.}
\label{fig7}
\end{figure}

We can obtain four-quark operators with a photon field by attaching a
photon to fermion lines in four-quark operators. First these operators
can be obtained from the annihilation and the $W$-exchange channels
from the four-quark operators in Eq.~(\ref{heff}) by attaching a
photon on each fermion except the spectator quark. The relevant
Feynman diagrams are shown in Fig.~\ref{fig7}. The Feynman diagrams in
Fig.~\ref{fig7} give the desired operators in
$\mathrm{SCET}_{\mathrm{I}}$ by integrating out the off-shell
modes. But these are subleading since the effective 
operators involve an usoft quark. When we go down to
$\mathrm{SCET}_{\mathrm{II}}$, there are no enhancements and the
effect of these operators still remains subleading and we neglect
these operators at leading order in SCET. In Fig.~\ref{fig7}, the
Feynman diagram in which a photon is attached to a 
spectator quark is missing, but it has been considered in the previous
subsection in which we consider the four-quark operators of the
form $(\overline{\chi}\Gamma_1   h)(\overline{\xi} \Gamma_2 \xi)$ and
we calculate the time-ordered product of this operator with the
electromagnetic interaction. This is because the intermediate state
has momentum of order $m_b\lambda$, and it cannot be integrated out.

\begin{figure}[t]
\begin{center}
\epsfig{file=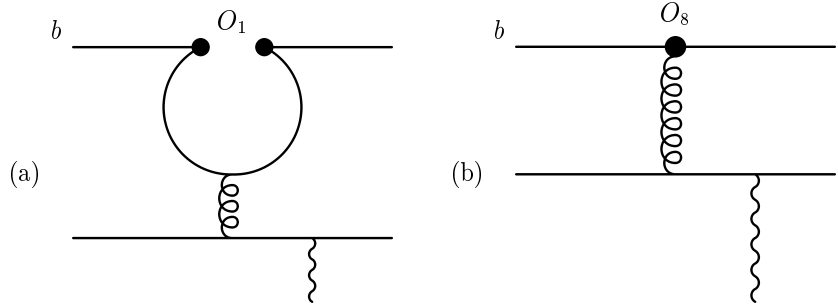, width=11.0cm}
\end{center}
\vspace{-0.5cm}
\caption{QCD diagrms to produce the four-quark operators with a
  photon arising from $O_1$ and $O_8$. These diagrams are subleading.}
\label{fig8}
\end{figure}

The effective four-quark operators with a photon can also arise from
the operators $O_1$ and $O_8$, 
as shown in Fig.~\ref{fig8}. However, the diagrams in Fig.~\ref{fig8}
are subleading due to the interaction of a collinear gluon with an
ultrasoft quark and a collinear quark. When we 
match this operator to $\mathrm{SCET}_{\mathrm{II}}$, there is no
enhancement, hence they are still subleading and neglected here. 

\section{Operators in $\mbox{\boldmath $\mathrm{SCET}_{\mathrm{II}}$
  \unboldmath}$  and their matrix  elements} 
\subsection{Form factors and the leading contributions}
We evaluate the matrix elements of all the relevant operators for
radiative $B$ decays, which are obtained in
$\mathrm{SCET}_{\mathrm{I}}$. And we decouple the usoft gluon
interactions by redefining collinear fields and go down to
$\mathrm{SCET}_{\mathrm{II}}$. There are various matrix
elements we should consider. First we evaluate matrix elements of the
operators without spectator quarks, namely $O_7^{(0)}$ and the
operators induced from $O_1$ and $O_8$, of which the matrix elements
are proportional to the matrix element of $O_7^{(0)}$. Second, we
consider the time-ordered products of the operators including external
gluons without spectator quarks
with the interaction of a collinear gluon with a collinear quark $\xi$
and a usoft quark. These matrix elements contribute to the form factor
$B\rightarrow V$. As will be seen later, some of the matrix elements
are factorized and they 
correspond to the contribution from the hard scattering amplitude,
which is calculable in perturbation theory. The remaining
nonfactorizable contributions involve endpoint singularities, but
these are absorbed into the soft form factor. The time-ordered
products from other operators, the time-ordered products of
four-quark operators with the electromagnetic interaction, and the
four-quark operators with a photon correspond to hard scattering
amplitudes with the convolution of the light-cone wave functions of
mesons. As discussed before, the contributions from the four-quark 
operators and the four-quark operators with a photon are subleading
and they will not be included in the final result for the decay
amplitudes at leading order in SCET.

In evaluating matrix elements of various contributions, the matrix
elements of heavy-to-light currents are related to form factors. The
form factors can be defined in SCET with a fewer number of indepedent
form factors due to the symmetry of SCET. In general, the form factors
for $B$ decays into vector mesons are defined as \cite{benfeld} 
\begin{eqnarray}
\langle V(p,\eta^*) |\overline{q} \gamma^{\mu} b |\overline{B} (p_B)
\rangle &=&\frac{2i V(q^2)}{m_B +m_V} \epsilon^{\mu\nu \alpha\beta}
\eta^*_{\nu} p_{\alpha} p_{B\beta}, \nonumber \\
\langle V(p,\eta^*) |\overline{q} \gamma^{\mu} \gamma_5 b
|\overline{B} (p_B) \rangle &=& 2m_V A_0 (q^2) \frac{\eta^* \cdot
  q}{q^2} q^{\mu} \nonumber \\
&& +(m_B + m_V) A_1 (q^2) \Bigl[ \eta^{*\mu}  -\frac{\eta^* \cdot
    q}{q^2} q^{\mu} \Bigr] \nonumber \\
&&-A_2 (q^2) \frac{\eta^* \cdot q}{m_B + m_V} \Bigl[p_B^{\mu} +
  p^{\mu} -\frac{m_B^2 -m_V^2}{q^2} q^{\mu} \Bigr], \nonumber \\
\langle V(p,\eta^*) |\overline{q} \sigma^{\mu\nu} q_{\nu} (1+\gamma_5)
b |\overline{B} (p_B) 
\rangle &=&2T_1 (q^2) \epsilon^{\mu\nu\alpha\beta} \eta^*_{\nu}
p_{B\alpha} p_{\beta} \nonumber \\
&&-iT_2 (q^2) \Bigl[ (m_B^2 -m_V^2) \eta^{*\mu} -(\eta^* \cdot q)
  (p_B^{\mu} + p^{\mu}) \Bigr] \nonumber \\
&&-iT_3 (q^2) (\eta^* \cdot q) \Bigl[ q^{\mu} -\frac{q^2}{m_B^2
    -m_V^2} (p_B^{\mu} +p^{\mu}) \Bigr], 
\label{fform}
\end{eqnarray}
where $q=p_B -p$, $m_V$ ($\eta$) is the mass (polarization vector) of
the vector meson and we use the convention $\epsilon^{0123} =-1$. Note
that the definiton of the form factors are different from that defined
by Ball and Braun \cite{ball}, but if we multiply the right-hand sides
of Eq.~(\ref{fform}) by the factor  $-i$, we obtain the definition in
Ref.~\cite{ball}.  

In SCET, we neglect the masses of the vector mesons, and the
polarization vector is transverse in this limit. We choose the plane
of the transverse polarization defined by $n\cdot \eta^* =
\overline{n} \cdot \eta^*=0$. In SCET, all the independent form
factors defined in Eq.~(\ref{fform}) are reduced to a few independent
nonperturbative functions. Especially, at leading order in SCET,
there is only one independent form factor $\zeta_{\perp}$, and the
matrix elements in Eq.~(\ref{fform}) reduce to
\cite{bauer2,chay1,charles} 
\begin{eqnarray}
\langle  V(n,\eta^*) | \overline{\xi} W \gamma^{\mu}  S^{\dagger} h
|\overline{B} \rangle &=& iE \zeta_{\perp} (E) \epsilon^{\mu\nu
  \alpha \beta} \eta^*_{\perp\nu} n_{\alpha} \overline{n}_{\beta},
\nonumber \\
\langle  V(n,\eta^*) | \overline{\xi} W \gamma^{\mu} \gamma_5
S^{\dagger} h |\overline{B} \rangle &=& 2E \zeta_{\perp} (E)
\eta^{*\mu}_{\perp}, \nonumber \\
\langle  V(n,\eta^*) | \overline{\xi} W \sigma^{\mu\nu} q_{\nu}
S^{\dagger} h |\overline{B} \rangle &=& -Em_B \zeta_{\perp} (E)
\epsilon^{\mu\nu \alpha\beta} \eta^*_{\perp\nu} n_{\alpha}
\overline{n}_{\beta}, \nonumber \\
\langle  V(n,\eta^*) | \overline{\xi} W \sigma^{\mu \nu} q_{\nu}
  \gamma_5  S^{\dagger} h |\overline{B} \rangle &=& -2iEm_B
\zeta_{\perp} (E) \eta^{*\mu}_{\perp}, 
\label{sform}
\end{eqnarray}
where the operators are replaced by the operators in
$\mathrm{SCET}_{\mathrm{II}}$ in a gauge invariant form. Note that
there is another nonperturbative function $\zeta_{\parallel}$, but
since we choose $n\cdot \eta_{\perp}^* = \overline{n} \cdot \eta^*=0$,
it does not appear in Eq.~(\ref{sform}) at leading order in SCET. 

From now on, we express all the operators at next-to-leading order
accuracy and replace the operators of the Wilson coefficients by
the Wilson coefficients after we apply the operators of the Wilson
coefficients inside the relevant operators. 
The leading-order contribution to the radiative $B$ decays comes from
the operator $O_7^{(0)}$, and in $\mathrm{SCET}_{\mathrm{II}}$, the
gauge invariant form under the collinear and soft gauge
transformations is given by
\begin{equation}
O_7^{(0)} = -\frac{em_b^2}{4\pi^2} \Bigl(\overline{\xi} W \Bigr)
\FMSlash{\mathcal{A}} (1-\gamma_5) S^{\dagger} h.
\end{equation}
The matrix element of $O_7^{(0)}$ is given by
\begin{equation}
\langle V(n,\eta_{\perp}), \gamma (\overline{n}, \epsilon_{\perp}) |
O_7^{(0)} |B \rangle = \frac{em_b^2}{8\pi^2} m_B
\zeta_{\perp} (E) 
\Bigl( 2\epsilon^*_{\perp} \cdot \eta^*_{\perp}
  +i\epsilon^{\mu\nu \alpha\beta}
\eta^*_{\perp\mu} \epsilon^*_{\perp\nu} n_{\alpha}
\overline{n}_{\beta} \Bigr),
\end{equation}
where $|B\rangle$ is an appropriate $B$ meson state depending on the
type of the vector meson $V$. 
 
There are also radiative corrections at order $\alpha_s$ from the
operators $O_1$ and $O_8$, in which there are an external photon and an
internal gluon. The matrix elements of
these operators are proportional to the matrix element of
$O_7^{(0)}$. Denoting the operators induced from $O_1$ and $O_8$ in
as $O_1^{(0)}$ and $O_8^{(0)}$, their matrix elements are
proportional to the matrix element of $O_7^{(0)}$, and can be
  written as \cite{bosch,hurth}
\begin{equation}
\langle O_{1,8}^{(0)}\rangle = \langle O_7^{(0)}\rangle \frac{\alpha_s
C_F}{4\pi} G_{1,8},
\end{equation}
where
\begin{eqnarray}
G_1 (s_c) &=& \frac{104}{27} \ln \frac{m_b}{\mu} +g_1 (s_c), \ G_8 =
-\frac{8}{3} \ln \frac{m_b}{\mu} +g_8, \nonumber \\
g_1 (s) &=& -\frac{833}{162} -\frac{20}{27} i\pi +\frac{8\pi^2}{9}
s^{3/2} \nonumber \\
&&+ \frac{2}{9} \Bigl[ 48 + 30i\pi -5\pi^2 -2i\pi^3 -36 \zeta(3) +(36
  +6i\pi -9\pi^2) \ln s \nonumber \\
&&+(3+6i\pi) \ln^2 s +\ln^3 s \Bigr]s \nonumber \\
&&+\frac{2}{9} \Bigl[ 18 +2\pi^2 -2i\pi^3 +(12-6\pi^2) \ln s + 6i\pi
  \ln^2 s +\ln^3 s \Bigr] s^2 \nonumber \\
&&+\frac{1}{27} \Bigl[ -9 + 112i\pi -14\pi^2 +(182-48i\pi) \ln s -126
  \ln^2 s \Bigr] s^3, \nonumber \\
g_8 &=& \frac{11}{3} -\frac{2\pi^2}{9} +\frac{2i\pi}{3},
\label{lcor}
\end{eqnarray}
and $s_c = m_c^2/m_b^2$. There can be subleading operators
$O_7^{(1a,1b)}$ from $O_1$ and $O_8$ in SCET, but they start with
$g^3$.

\subsection{Contributions to the form factor}
We consider first the effects of the operators $O_7^{(0,1a,1b)}$,
which are obtained from $O_7$. We calculate the
time-ordered products of these operators with the interaction of
collinear quarks with an usoft quark and a collinear quark $\xi$ in
$\mathrm{SCET}_{\mathrm{I}}$. The interaction of collinear and usoft
gluons starts from order $\lambda$, but since the propagator of the
exchanged gluon is enhanced by $1/\lambda^2$, we have to include the
interaction of collinear and usoft gluons to order $\lambda^2$ to
obtain the leading-order result. Among these contributions, the
time-ordered products of the operator derived from $O_7$ contribute to
the form factor, while those from the operators $O_1$ and $O_8$
contribute to the hard scattering amplitudes. Here we first consider
the  contributions to the form factor.

\begin{figure}[t]
\begin{center}
\epsfig{file=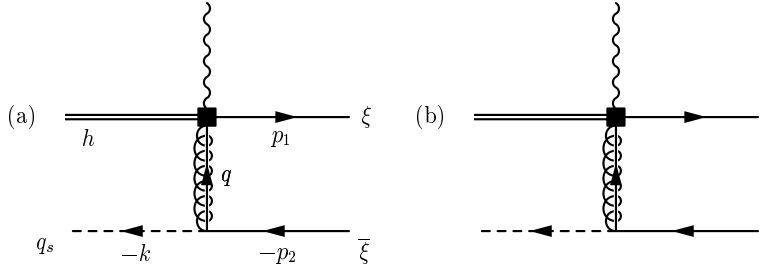, width=10.0cm}
\end{center}
\caption{Tree-level graphs in $\mathrm{SCET}_{\mathrm{I}}$ for the
  spectator contribution to the heavy-to-light form factor. The first
  diagram contributes to $T_{1,3}$, and the second diagram
  contributes to $T_{0,2,3,4,5}$.}
\label{fig9}
\end{figure}

The Lagrangian for the interactions of collinear gluons with a
collinear quark $\xi$ and an usoft quark $q_{us}$ is given by
\cite{beneke1,bs,bauer5}  
\begin{eqnarray} 
\mathcal{L}_{\xi q}^{(1)} &=& ig \overline{\xi}
\frac{1}{i\overline{n}\cdot D_n} \FMSlash{B}_{\perp}^n W
  q_{us} +\mathrm{h. c.}, \ \mathcal{L}_{\xi q}^{(2a)} =ig
  \overline{\xi} \frac{1}{i\overline{n} \cdot D_n} \FMSlash{M}
  Wq_{us} +\mathrm{h.c.}, \nonumber \\
\mathcal{L}_{\xi q}^{(2b)} &=& ig \overline{\xi} \frac{\FMslash{n}}{2}
i\FMSlash{D}_{\perp}^n \frac{1}{(i\overline{n} \cdot D_n)^2}
\FMSlash{B}_{\perp}^n W q_{us} +\mathrm{h.c.},
\end{eqnarray}
where
\begin{equation}
ig \FMSlash{B}_{\perp}^n = [i\overline{n}\cdot D_n,
  i\FMSlash{D}_{\perp}^n], \ ig \FMSlash{M} =[i\overline{n} \cdot D^n,
  i\FMSlash{D}^{us} +\frac{\FMslash{\overline{n}}}{2} gn\cdot
  A_n].
\end{equation}
At leading order in $\mathrm{SCET}_{\mathrm{I}}$, the time-ordered
products, which contribute to the form factor, are given as
\begin{eqnarray}
T_0^F &=&\int d^4 x T[O_7^{(0)} (0) i\mathcal{L}_{\xi q}^{(1)} (x) ], \
T_1^F = \int d^4 x T[O_7^{(1)} (0) i\mathcal{L}_{\xi q}^{(1)} (x)],
 \\ 
T_2^F &=&
\int d^4 x T[O_7^{(0)} (0) i\mathcal{L}_{\xi q}^{(2b)} (x) ], \ 
T_3^{NF} =
\int d^4 x T[O_7^{(0)} (0) i\mathcal{L}_{\xi q}^{(2a)} (x) ], \nonumber
\\  
T_4^{NF}
&=& \int d^4 x d^4 y T[O_7^{(0)} (0) i\mathcal{L}_{\xi \xi}^{(1)} (x)
  i\mathcal{L}_{\xi q}^{(1)} (y) ], \
T_5^{NF}
= \int d^4 x d^4 y T[O_7^{(0)} (0) i\mathcal{L}_{cg}^{(1)} (x)
  i\mathcal{L}_{\xi q}^{(1)} (y) ], \nonumber 
\label{formtime}
\end{eqnarray}
and these are shown in Fig.~\ref{fig9} schematically.
After we evaluate the matrix elements of the time-ordered products in
$\mathrm{SCET}_{\mathrm{I}}$, we decouple the
collinear-usoft interaction using the field redefinitions 
\begin{equation}
\xi^{(0)} = Y^{\dagger} \xi, \ \ A_n^{(0)} = Y^{\dagger}A_n Y,
\ \ 
Y(x) = \mathrm{P} \exp \Bigl( ig \int_{-\infty}^x ds n\cdot
  A_{us} (ns) \Bigr),
\label{softd}
\end{equation}
in order to go down to $\mathrm{SCET}_{\mathrm{II}}$.
Matching at $\mu_0 \sim \sqrt{m_b \Lambda}$, the usoft fields become
soft ($Y\rightarrow S$) and the operators are matched onto the
operators in $\mathrm{SCET}_{\mathrm{II}}$. When we redefine the
fields using the Wilson line $Y$, soft gluons decouple in $T_i^F$. It
means that the contributions are factorizable to all orders in
$\alpha_s$. On the other hand, soft gluons do not decouple in the
time-ordered products $T_i^{NF}$. Furthermore these contributions have
endpoint singularities \cite{bs}. They are absorbed in the soft form
factor $\zeta_{\perp}$, as was done in Ref.~\cite{bs}.

Among the factorizable time-ordered products, $\langle T_0^F\rangle
=\langle T_2^F\rangle =0$. The matrix element of $T_1^F$ is given by
\begin{eqnarray}
\langle T_1^F\rangle &=& \langle V, \gamma|\int d^4 x T[O_7^{(1)} (0)
  i\mathcal{L}_{\xi q}^{(1)} (x)] |B\rangle \nonumber \\
&=& \frac{\alpha_s}{\pi} \frac{C_F}{N} \frac{em_b}{16\pi} \int
  d\overline{n} \cdot x \int dn\cdot k  e^{in\cdot k
  \overline{n} \cdot x/2} \nonumber \\
&&\times \langle V,\gamma| (\overline{q}_s S)(\overline{n}\cdot x)
  \frac{1}{n\cdot \mathcal{R}^{\dagger}}
  \FMslash{n} (1+\gamma_5) S^{\dagger} h \cdot \overline{\xi} W
  \frac{1}{\overline{n}\cdot \mathcal{P}}
  \FMSlash{\mathcal{A}} \FMslash{\overline{n}} (1+\gamma_5) W
  \xi|B\rangle\nonumber \\
&=& \frac{e\alpha_s}{8\pi} \frac{C_F}{N}  f_B f_{\perp} m_B^2 \Bigl(
  2\eta_{\perp}^* \cdot \epsilon_{\perp}^* +i\epsilon_{\mu\nu
  \alpha\beta} \eta_{\perp}^{*\mu} \epsilon_{\perp}^{*\nu} n^{\alpha}
  \overline{n}^{\beta} \Bigr) \int dr_+ \frac{\phi_B^+ (r_+)}{r_+}
  \int du \frac{\phi_{\perp} (u)}{\overline{u}},
\label{t1f}
\end{eqnarray}
where $u$ ($\overline{u} \equiv 1-u$) is the momentum fraction of the
quark (antiquark) inside the meson. $\mathcal{R}$ is the operator
which extracts a soft momentum (of order $\Lambda_{\mathrm{QCD}}$)
from a soft particle. In evaluating Eq.~(\ref{t1f}), the  
matrix element involving the $B$ meson can be calculated as
\begin{eqnarray}
&&\langle 0| \overline{q}_s S (\overline{n} \cdot x) \FMslash{n}
(1+\gamma_5) S^{\dagger} h |B\rangle  \nonumber \\
&&=\int dr_+ e^{-ir_+ \overline{n} \cdot x/2} \mathrm{Tr} \ \Bigl[
    \Psi_B (r_+) \FMslash{n} (1+\gamma_5) \Bigr] \nonumber \\
&&=-\frac{if_B m_B}{4} \int dr_+  e^{-ir_+ \overline{n} \cdot x/2}
  \mathrm{Tr}\ \Bigl[ \frac{1+\FMslash{v}}{2} \FMslash{\overline{n}}
    \gamma_5 \FMslash{n} (1+\gamma_5) \Bigr] \phi_B^+ (r_+) \nonumber
  \\
&&= if_B m_B \int dr_+  e^{-ir_+ \overline{n} \cdot x/2}
  \phi_B^+ (r_+),
\end{eqnarray}
where the leading-twist $B$ meson light-cone wave function is defined
through the projection of the $B$ meson as \cite{benfeld,grozin}
\begin{equation}
\Psi_B (r_+) =-\frac{if_B m_B}{4} \Bigl[ \frac{1+\FMslash{v}}{2}
    \Bigl( \FMslash{\overline{n}} \phi_B^+ (r_+) + \FMslash{n}
    \phi_B^- (r_+) \Bigr) \gamma_5 \Bigr].
\end{equation}

And the momentum-space representation of the vector meson light-cone
projection at leading order is given by
\begin{equation}
M_{\alpha \beta}^V = M_{\alpha\beta \parallel}^V +
M_{\alpha\beta\perp}^V,
\end{equation}
where, at leading twist and at leading order in SCET with $n\cdot
\eta_{\perp}^* = \overline{n} \cdot \eta_{\perp}^* =0$ 
\begin{equation}
M_{\parallel}^V = 0, \ \ M_{\perp}^V = -\frac{if_{\perp}}{4} E
\FMslash{\eta}_{\perp}^*  \FMslash{n} \phi_{\perp} (u),
\end{equation}
where $\eta_{\perp}^{*\mu}$ is the transverse polarization of the
vector meson.

Among the nonfactorizable contributions, $\langle T_5^{NF}\rangle
=0$. In $\langle T_{3,4}^{NF}\rangle$, there exist $1/u^2$, and
$1/r_+^2$ singularities. However, since all endpoint singularities are
regulated by $\Lambda_{\mathrm{QCD}}$ in the full QCD, if all the soft
operators are included to cover the endpoint regions, the
singularities will not arise. We absorb the nonfactorizable
contributions to the form factor into the soft form factor, given by
$\zeta_{\perp}$. 

\subsection{Contributions to hard scattering amplitudes}
There are other time-ordered products of the operators derived from
$O_{1,8}$, which contribute to the nonfactorizable contributions to
the radiative decays, which are defined as 
\begin{equation}
U_1 =
\int d^4 x T[O_1^{(1)} (0) i\mathcal{L}_{\xi q}^{(1)} (x) ], \ 
U_2 =
\int d^4 x T[O_8^{(1)} (0) i\mathcal{L}_{\xi q}^{(1)} (x) ], 
\end{equation}
where the operators $O_{1,8}^{(1)}$ are the effective operators
derived from $O_{1,8}$ at order $\lambda$. These operators are given as
\begin{eqnarray}
O_1^{(1)} &=&Q_u A_1^{(1a)} O_7^{(1a)} +Q_u A_1^{(1b)} O_7^{(1b)}
  \nonumber \\
&=&\frac{eQ_u m_b}{16\pi^2}  \Bigl(\overline{\xi} W
H(\overline{n} \cdot \mathcal{P}, s_i) 
  \FMSlash{\mathcal{A}} [W^{\dagger} i\FMSlash{D}_{n\perp} W]
  (1+\gamma_5) h \nonumber \\
&&-\overline{\xi} W H(\overline{n} \cdot \mathcal{P}, s_i) 
 [W^{\dagger}   i\FMSlash{D}_{n\perp} W] 
  \FMSlash{\mathcal{A}} (1+\gamma_5) h \Bigr), \nonumber \\
O_8^{(1)} &=& Q_d A_8^{(1a)} O_7^{(1a)} \nonumber \\
&=&
\frac{eQ_d m_b}{4\pi^2} \overline{\xi} W
\FMSlash{\mathcal{A}} \frac{1}{\overline{n} \cdot
  \mathcal{P}^{\dagger}} \overline{n} \cdot \mathcal{P}
  [W^{\dagger}i\FMSlash{D}_{n\perp} W] 
  (1+\gamma_5) h,
\label{o181}
\end{eqnarray}
which can be obtained from Eqs.~(\ref{o1gp}) and (\ref{o8gp}). Since
the operators $O_{1,8}^{(1)}$ start with $\lambda$, we need
time-ordered products only with $\mathcal{L}_{\xi q}^{(1)}$.

The matrix element of $U_1$ is written as
\begin{eqnarray}
\langle U_1\rangle &=&\langle V, \gamma| \int  d^4 x T[O_1^{(1)} (0)
  i\mathcal{L}_{\xi q}^{(1)} (x) ] |B\rangle  \nonumber \\
&=& \frac{\alpha_s}{4\pi} \frac{C_F}{N} \frac{eQ_u}{16\pi} \int
  d\overline{n} \cdot x \int dn\cdot k e^{i n\cdot k \overline{n}
  \cdot x/2} \nonumber \\
&&\times \langle V,\gamma| (\overline{q}_s S)(\overline{n} \cdot x)
  \frac{1}{n \cdot \mathcal{R}^{\dagger}} \FMslash{n} (1+\gamma_5)
  S^{\dagger} h  \cdot \overline{\xi} WH(\overline{n} \cdot \mathcal{P},
  s_i) \FMSlash{\mathcal{A}} \frac{1}{\overline{n} \cdot
  \mathcal{P}}  \FMslash{\overline{n}}
  (1+\gamma_5) W^{\dagger} \xi |B\rangle \nonumber \\
&=& \frac{\alpha_s}{32\pi} \frac{C_F}{N} eQ_u f_B
f_{\perp} m_B^2 \Bigl(
  2\eta_{\perp}^* \cdot \epsilon_{\perp}^* +i\epsilon_{\mu\nu
  \alpha\beta} \eta_{\perp}^{*\mu} \epsilon_{\perp}^{*\nu} n^{\alpha}
  \overline{n}^{\beta} \Bigr)\nonumber \\
&&\times \int dr_+ \frac{\phi_B^+ (r_+)}{r_+}
  \int du H(m_b u,s_i) \frac{\phi_{\perp}
  (u)}{\overline{u}}. \label{t3f}  
\end{eqnarray}
If we use the leading-twist light-cone wave function $\phi_{\perp} (u)
= 6u\overline{u}$ for the vector meson, we can replace $\overline{u}$
by $u$ in the last line of Eq.~(\ref{t3f}). But in general, if the
valence quarks have different masses such that the wave function is not
symmetric under $u\leftrightarrow \overline{u}$, the present form should
be kept. 

Note that the matrix element of the time-ordered product with the
second operator of $O_1^{(1)}$ in Eq.~(\ref{o181}), or the operator
$O_7^{(1b)}$ in Eq.~(\ref{o7op2}), vanishes. To prove this, let us
consider the matrix element of the operator
\begin{equation}
O_7^{(1b)}= \frac{em_b}{4\pi^2}\overline{\xi} W  [W^{\dagger}
  i\FMSlash{D}_{n\perp} W]  
  \FMSlash{\mathcal{A}} (1+\gamma_5)  h.
\end{equation}
The Wilson coefficients, when
sandwiched between the heavy quark and the collinear quark fields,
produce kinematic variables such as $\overline{n}\cdot p_1$, where
$p_1$ is the momentum of the collinear quark $\xi$. Then the matrix
element of the operator $O_7^{(1b)}$ with the collinear Lagrangian
$\mathcal{L}_{\xi q}^{(1)}$ is written as
\begin{eqnarray}
&&\langle V, \gamma|\int d^4 x T[O_7^{(1b)}(0) i\mathcal{L}_{\xi
      q}^{(1)} (x) 
]|B\rangle = \frac{\alpha_s C_F}{4N} \frac{em_b}{4\pi^2} \int
      d\overline{n} \cdot x \int 
dn\cdot k e^{in\cdot k \overline{n} \cdot x/2}\nonumber \\
&&\times \langle (\overline{q}_s S)(\overline{n} \cdot x)
\frac{1}{n\cdot \mathcal{R}^{\dagger}} \FMslash{n}
\FMSlash{\mathcal{A}} (1+ \gamma_5) S^{\dagger} h \cdot \overline{\xi}
W \frac{1}{\overline{n} \cdot \mathcal{P}} \overline{\FMslash{n}}
(1-\gamma_5) W^{\dagger} \xi |B \rangle
\label{too} 
 \end{eqnarray}
The matrix element in the last line in Eq.~(\ref{too}) is zero when we
use the $B$ mesons and the vector meson projections at leading
order. Therefore the matrix element of the time-ordered products of
$O_7^{(1b)}$ with $\mathcal{L}_{\xi q}^{(1)}$ vanishes.

Similarly the matrix element of $U_2$ is written as
\begin{eqnarray}
\langle U_2\rangle &=& \langle V,\gamma| \int d^4 x T[O_8^{(1)} (0)
  i\mathcal{L}_{\xi q}^{(1)} (x) ] |B\rangle \label{t2f}
\\
&=& \frac{\alpha_s}{\pi} \frac{C_F}{N} \frac{eQ_d m_b}{16\pi} \int
  d\overline{n} \cdot x \int dn\cdot k e^{i n\cdot k \overline{n}
  \cdot x/2} \nonumber \\
&&\times \langle V,\gamma| (\overline{q}_s S)(\overline{n} \cdot x)
  \frac{1}{n \cdot \mathcal{R}^{\dagger}} \FMslash{n} (1+\gamma_5)
  S^{\dagger} h  \cdot \overline{\xi} W \frac{1}{\overline{n} \cdot
  \mathcal{P}^{\dagger}} \FMSlash{\mathcal{A}} \FMslash{\overline{n}}
  (1+\gamma_5) W^{\dagger} \xi |B\rangle \nonumber \\
&=& -eQ_d \frac{\alpha_s}{8\pi} \frac{C_F}{N} f_B f_{\perp} m_B^2
  \Bigl( 2\eta_{\perp}^* \cdot \epsilon_{\perp}^* +i\epsilon_{\mu\nu
  \alpha\beta} \eta_{\perp}^{*\mu} \epsilon_{\perp}^{*\nu} n^{\alpha}
  \overline{n}^{\beta} \Bigr) \int dr_+ \frac{\phi_B^+ (r_+)}{r_+}
  \int du \frac{\phi_{\perp} (u)}{u}. \nonumber 
\end{eqnarray}

\subsection{Contributions of the operators $(\overline{\chi}\Gamma_1
  h)(\overline{\xi} \Gamma_2 \xi)$ }
In $\mathrm{SCET}_{\mathrm{I}}$, the contributions of the operators
$(\overline{\chi}\Gamma_1   h)(\overline{\xi} \Gamma_2 \xi)$ can be
calculated through the time-ordered product with the electromagnetic
current interacting with an usoft quark and a collinear quark. The
electromagnetic interaction with an usoft quark and a collinear quark
is given by 
\begin{equation}
\mathcal{L}^{\mathrm{em}}_{\chi q}  = eQ_{sp} \overline{q}_{us}
\FMSlash{\mathcal{A}} \overline{W}^{\dagger} \chi,
\label{emint}
\end{equation}
where $eQ_{sp}$ is the electric charge of the spectator quark. To be
complete, we need to include a nonlocal term and the electromagnetic
interaction should read
\begin{equation}
\overline{q} \FMSlash{\mathcal{A}} q \rightarrow \overline{q}_{us}
\FMSlash{\mathcal{A}} \overline{W}^{\dagger} \chi +T[\mathcal{T}_g,
  i\mathcal{L}_{\xi q}^{(1)}],
\end{equation}
where the operators in the nonlocal term are defined as
\begin{eqnarray}
\mathcal{T}_g &=&\overline{\chi} \Bigl[\FMSlash{\mathcal{A}}
  \frac{1}{n\cdot (\mathcal{Q}+gA_{\overline{n}})}
  (\FMSlash{\mathcal{Q}}_{\perp} +g\FMSlash{A}_{\overline{n} \perp})
+ (\FMSlash{\mathcal{Q}}_{\perp} +g\FMSlash{A}_{\overline{n} \perp})
  \frac{1}{n\cdot (\mathcal{Q}+gA_{\overline{n}})}
  \FMSlash{\mathcal{A}} \Bigr] 
  \frac{\FMslash{n}}{2} \chi, \nonumber \\
\mathcal{L}_{\chi q}^{(1)} &=& \overline{q}_{us}
  \overline{W}^{\dagger} (\FMSlash{\mathcal{Q}}_{\perp}
  +g\FMSlash{A}_{\overline{n} \perp} ) \chi +\mathrm{h.c.}.
\label{tem}
\end{eqnarray}
The operators in Eq.~(\ref{tem}) are
the same as those defined in Ref.~\cite{lunghi} except that $n^{\mu}$
and $\overline{n}^{\mu}$ are interchanged. The matching of the
operators to the full theory, and the computation of the Wilson
coefficients have been extensively discussed in
Ref.~\cite{lunghi}. But the main point here is that the 
electromagnetic interaction  at leading
order in SCET has the same operator structure as given in
Eq.~(\ref{emint}), but with a different Wilson coefficient including
all the effects with a nonlocal term. We will show that the
contribution of the time-ordered 
products of the four-quark operators with the electromagnetic
interaction vanishes at leading order in SCET, and we only need the
form of the operator in the argument.
 
All the contributions from the four-quark operators through the
annihilation channels and the $W$-exchange channels vanish at leading
order. The proof is the following. Neglecting the color indices, the
four-quark operators can be generically written as
\begin{equation}
Q_i = (\overline{\chi} \overline{W} \Gamma_1  h) (
\overline{\xi} W \Gamma_2 W^{\dagger} \xi), 
\end{equation}
where $\Gamma_1$, $\Gamma_2$ are Dirac matrices. 
\begin{figure}[t]
\begin{center}
\epsfig{file=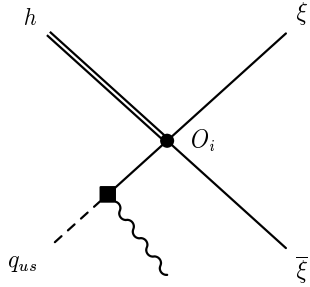, width=4.0cm}
\end{center}
\caption{Time-ordered products of the four-quark operators with the
  electromagnetic interaction of an usoft quark and a collinear quark
  $\chi$. The circle represents the
  four-quark operators, and the square
  represents the electromagnetic interaction.}
\label{fig10}
\end{figure}
The time-ordered products with the electromagnetic current, which is
schematically shown in Fig.~\ref{fig10},  are given by
\begin{eqnarray}
\langle V_i\rangle &=& \langle V,\gamma| \int d^4 x T[Q_i (0),
  i\mathcal{L}^{em}_{q\chi} (x)] |B\rangle \nonumber \\
&=&-\frac{eQ_{sp}}{8\pi} \int dn\cdot x \int \frac{d \overline{n} \cdot
  k}{\overline{n} \cdot k} e^{i\overline{n} \cdot k n\cdot x/2}
  \langle V,\gamma| \overline{\xi} \Gamma_2 \xi \cdot
  \overline{q}_s \FMSlash{\mathcal{A}} \FMslash{\overline{n}} \Gamma_1
  h |B \rangle =0.
\end{eqnarray}
The matrix element vanishes at leading order in SCET when we take the
matrix element of the collinear sector $\overline{\xi} \Gamma_2 \xi$
for $\Gamma_2 = \FMslash{\overline{n}} (1\pm \gamma_5)$, or $1\pm
\gamma_5$ with the vector meson, they vanish using the leading
projection for the vector meson. Therefore at leading order in SCET,
the annihilation channels and the $W$-exchange channels do not
contribute to the radiative $B$ decays, though there may be subleading 
contributions. For example, the isopin breaking effects in
$B\rightarrow V\gamma$ decays are explained by the effects of these
operators.  

\section{Decay amplitudes for $\mbox{\boldmath $B\rightarrow V\gamma$
    \unboldmath}$ }
Combining all the contributions considered in the previous section, we
can write the decay amplitudes for radiative $B$ decays. In general,
the decay amplitudes can be written as
\begin{eqnarray}
A(B\rightarrow V\gamma) &=& \frac{G_F}{\sqrt{2}} \frac{em_b
  m_B^2}{8\pi^2} c_V \Bigl(2\epsilon_{\perp}^* \cdot \eta_{\perp}^*
+i\epsilon^{\mu \nu\alpha\beta} \eta^*_{\perp\mu} \epsilon^*_{\perp
  \nu} n_{\alpha} \overline{n}_{\beta} \Bigr) \sum_{p=u,c} V_{pd}^*
V_{pb}\nonumber \\
&&\times \Bigl[ C_7 (m_b) A_7^{(0)} (\mu_0) \frac{m_b}{m_B}
  \zeta_{\perp} (E,\mu) + 
  \frac{\alpha_s  C_F}{4\pi} C_i (m_b) G_i (m_b) A_i (\mu_0)
  \frac{m_b}{m_B} \zeta_{\perp}    (E,\mu) \nonumber \\
&&+\frac{f_B f_{\perp}}{m_b} \int du dr_+ T^a (\mu_0) J_a (u,r_+,
  \mu_0,\mu)  \phi_B^+ (r_+,\mu) \phi_{\perp} (u,\mu) \Bigr],
\end{eqnarray}
where $c_V=1$ for $V=K^*$, $\rho$, and $c_V=1/\sqrt{2}$ for
$V=\rho^0$, and $i=1,8$. $C_i (m_b)$ are the Wilson coefficients in
the full theory, evaluated at $\mu=m_b$, and $A_i (\mu_0)$ are the
Wilson coefficients in $\mathrm{SCET}_{\mathrm{I}}$, evaluated at $\mu_0 =
\sqrt{m_b \Lambda_{\mathrm{QCD}}}$. $G_i$ are the loop
corrections, which are given in Eq.~(\ref{lcor}) at order
$\alpha_s$. The functions $T_a (\mu_0)$ are  
the products of the Wilson coefficients in the full QCD at $\mu=m_b$,
and the Wilson coefficients in $\mathrm{SCET}_{\mathrm{I}}$ at
$\mu=\mu_0$. The functions $J_a$ are the jet functions which can be
obtained in matching $\mathrm{SCET}_{\mathrm{I}}$ and
$\mathrm{SCET}_{\mathrm{II}}$. 

The decay amplitudes 
for radiative $B$ decays  at leading order in SCET and at
next-to-leading order in $\alpha_s$ are written as
\begin{eqnarray}
A (B \rightarrow V \gamma) &=& \frac{G_F}{\sqrt{2}}c_V \sum_{p=u,c}
 V_{pd} V_{pb}^*  \Bigl[
 \Bigl(C_7 (m_b) A_7^{(0)} (\mu_0) +\frac{\alpha_s C_F}{4\pi} (C_1 G_1
 +C_8  G_8) \Bigr)  \langle V,\gamma | O_7^{(0)} |B\rangle \nonumber
 \\
&& +C_7 (m_b) \langle T_1^F \rangle  
+ C_1 (m_b) \langle U_1 \rangle + C_8 (m_b) \langle U_2 \rangle
 \Bigr] \nonumber \\
&=& \frac{G_F}{\sqrt{2}} \frac{em_b}{8\pi^2}
m_B^2 c_V  \Bigl(2\epsilon_{\perp}^* \cdot \eta_{\perp}^*
+i\epsilon^{\mu \nu\alpha\beta} \eta^*_{\perp\mu} \epsilon^*_{\perp
  \nu} n_{\alpha} \overline{n}_{\beta} \Bigr) \sum_{p=u,c} V_{pd}^*
V_{pb}\nonumber \\
&\times& \Bigl[ C_7 (m_b) \Bigl( A_7^{(0)}
  (\mu_0) \frac{m_b}{m_B} \zeta_{\perp} (E) +\pi \alpha_s
 \frac{C_F}{N} \frac{f_B 
    f_{\perp}}{m_b m_B} m_B\int dr_+ \frac{\phi_B^+ (r_+)}{r_+} \int
  du \frac{\phi_{\perp} (u)}{\overline{u}} \Bigr) \nonumber \\
&&+ C_1 (m_b) \Bigl( \frac{\alpha_s C_F}{4\pi} \frac{m_b}{m_B}
 \zeta_{\perp} (E) G_1   (s_p) \nonumber \\
&& +\frac{\pi \alpha_s}{6} \frac{C_F}{N} \frac{f_B
f_{\perp}}{m_b m_B} m_B \int dr_+ \frac{\phi_B^+ (r_+)}{r_+}
  \int du H(m_b u,s_i) \frac{\phi_{\perp} (u)}{\overline{u}} \Bigr)
  \nonumber \\
&&+C_8 (m_b) \Bigl( \frac{\alpha_s C_F}{4\pi} \frac{m_b}{m_B}
 \zeta_{\perp} (E) G_8 
(s_p) \nonumber \\
&&+ \frac{\pi\alpha_s}{3} \frac{C_F}{N} \frac{f_B f_{\perp}}{m_b m_B} 
m_B\int dr_+ \frac{\phi_B^+ (r_+)}{r_+}
  \int du \frac{\phi_{\perp} (u)}{u} \Bigr) \Bigr].
\label{damp}
\end{eqnarray}
The Wilson coefficient $A_7^{(0)}$ to next-to-leading order accuracy is
given by Eq.~(\ref{a70}), and all the other Wilson coefficients are 1
at this order because the Wilson coefficients appearing in the terms
other than the first term in Eq.~(\ref{damp}) is proportional to
$\alpha_s$. In fact we have to use the evolution of the Wilson
coefficient $A_7^{(0)}(\mu)$ from $m_b$ to $\mu_0$ using the
renormalization group equation because there is a double
logarithm. And the evolution of other Wilson coefficients should be
performed, which is not available yet. However, since the evolution
from $\mu=m_b$ to $\mu = \mu_0$ is small, the numerical difference
would be small.  

In the final expression of Eq.~(\ref{damp}), the first term
proportional to $C_7$ is the contribution to
the form factor $B\rightarrow V$. The nonfactorizable part $\langle
T_i^{NF} \rangle$ has been
absorbed in the soft form factor $\zeta_{\perp}$, and the hard
scattering contribution can be calculated. The form factor $F_V$ for
$B \rightarrow V$ as
\begin{eqnarray}
\langle V(n,\eta_{\perp}) \gamma (\overline{n},\epsilon_{\perp})
|O_7 |B\rangle &=& \frac{em_b}{8\pi^2} m_B^2 c_V
F_V \nonumber \\
&&\times \Bigl(2\epsilon_{\perp}^* \cdot \eta_{\perp}^*
+i\epsilon^{\mu \nu\alpha\beta} \eta^*_{\perp\mu} \epsilon^*_{\perp
  \nu} n_{\alpha} \overline{n}_{\beta} \Bigr), 
\end{eqnarray}
where the $\overline{B}\rightarrow V$ form factor $F_V$ is
evaluated at $q^2=0$, and in terms of the nonperturbative function
$\zeta_{\perp}$, it is written, at leading order in SCET, as
\begin{equation}
F_V = A_7^{(0)} (\mu_0) \frac{m_b}{m_B} \zeta_{\perp} (E,\mu) + \pi
\alpha_s (\mu) \frac{C_F}{N} \frac{f_B f_{\perp}}{m_b m_B} m_B \int
dr_+ \frac{\phi_B^+ (r_+,\mu)}{r_+} \int du \frac{\phi_{\perp}
  (u,\mu)}{\overline{u}},  
\label{fv}
\end{equation}
where nonfactorizable contributions to the form factor from
$T_i^{NF}$, which include the singularities in $1/u^2$ and $1/r_+^2$ 
are absorbed to the soft form factor $\zeta_{\perp}$ \cite{bs}. 
This result is consistent with the result of Ref.~\cite{bosch}, which
was obtained at leading order in $\Lambda_{\mathrm{QCD}}/m_b$ in the
heavy quark limit. Therefore we conclude that the leading-order result
in SCET corresponds to the leading-order result in the heavy quark
mass limit, as it should be. 

In Ref.~\cite{bosch}, they consider the scale dependence of the matrix
element of $O_7$ by including the running of the product of the $b$
quark mass and form factor as
\begin{equation}
(m_b F_V) [\mu] = (m_b F_V)[m_b] \Bigl(1+\frac{\alpha_s}{4\pi} 8C_F
  \ln \frac{m_b}{\mu} \Bigr),
\end{equation}
but in our formalism, we use the running $b$ quark mass at $\mu=m_b$
and the Wilson coefficient $A_7^{(0)}$ evaluated at $\mu_0$, and the
  scaling of $F_V$ is manifestly expressed in Eq.~(\ref{fv}). 

\section{Conclusion}
We can apply SCET to the radiative $B$ decays such as $\overline{B}
\rightarrow K^* \gamma$ or $\overline{B} \rightarrow \rho \gamma$, and
the organization of the relevant operators can be systematically
achieved using the power counting in SCET. Employing the two-step
matching from $\mathrm{SCET}_{\mathrm{I}}$ to
$\mathrm{SCET}_{\mathrm{II}}$, off-shell modes are integrated out
successively and we can obtain gauge invariant operators under
collinear and soft gauge transformations in powers of $\lambda$. The
scaling of $\lambda$ changes from $\sqrt{\Lambda_{\mathrm{QCD}}/m_b}$
in $\mathrm{SCET}_{\mathrm{I}}$ to $\Lambda_{\mathrm{QCD}}/m_b$ in
$\mathrm{SCET}_{\mathrm{II}}$, and the power counting of the
operators can be achieved including the time-ordered products of the
operators with the Lagrangian. The matrix elements of the operators
can be evaluated in $\mathrm{SCET}_{\mathrm{II}}$ including the
effects of  the evolution of the operators. We considered four-quark
operators through the annihilation channels and the $W$ exchange
channels in SCET, and all these effects turn out to be at subleading
order by our power counting method.

All the hard scattering amplitudes are shown to be factorized and the
decay amplitudes can be written as the convolution of the hard
scattering amplitudes with the light-cone wave functions of the
mesons. For the contribution to the form factor, we can categorize the
time-ordered products of $O_7^{(0,1a)}$ with the interaction of a
collinear gluon with a collinear quark $\xi$ and a soft quark into
factorizable and nonfactorizable contributions. The factorizable
contributions are calculable as a convolution of the hard scattering
amplitude and light-cone wave functions of mesons. In the nonfactorizable
contributions to the form factor, the effects of soft gluons are not
decoupled and they are purely nonperturbative effects, and they have
endpoint $1/u^2$, $1/r_+^2$ singularities. These
nonfactorizable contributions to the form factor are absorbed into the
definition of the nonperturbative form factor $\zeta_{\perp}$.

We have explicitly shown that the decay amplitudes are factorized to
all orders in $\alpha_s$ at leading order in SCET. This is an
extension of the factorization theorem proved at next-to-leading order
in the heavy quark limit \cite{bosch}. It is also shown that the decay
amplitudes at leading order in SCET coincide with the amplitudes
obtained in the heavy quark limit at leading order in
$\Lambda_{\mathrm{QCD}}/m_b$. This should be true because the two
limits at leading order are the same, and this correspondence was also
shown in nonleptonic decays into two light mesons \cite{chay2,chay3}. We
have calculated the decay amplitudes at next-to-leading order. But
this does not mean that the factorization theorem is valid only to this
order. The operators obtained in SCET are gauge invariant by attaching
the Wilson lines, and these include the effects of gluons to all
orders. If we analyze radiative $B$ decays at next-to-next-to-leading
order, the only changes occur in the Wilson coefficients and we have
to include more time-ordered products, the form of the operators does
not change. And the factorization properties remain intact. Therefore
the factorization theorem works to all orders in 
$\alpha_s$ at leading order in SCET, and we have evaluated the Wilson
coefficients at next-to-leading order in this paper.

The contributions of the annihilation channels and the $W$-exchange
channels are subleading. This point was also pointed out in the heavy
quark mass limit \cite{bosch}. Therefore the isospin breaking effects
due to the different contents of the spectator quarks in the $B$ meson
appear only at subleading order in $\Lambda_{\mathrm{QCD}}/m_b$ in the
heavy quark limit \cite{kagan}. The $SU(3)$ breaking effects which
arise in the difference between the decay rates
$\overline{B}\rightarrow K^* 
\gamma$ and $\overline{B}\rightarrow \rho \gamma$ also appear at
subleading order. It would be challenging to consider radiative $B$
decays to order $\Lambda_{\mathrm{QCD}}$ in SCET. In order to probe
subleading effects such as the isospin breaking effects and the
$SU(3)$ flavor breaking effects, we have to devise a systematic method
to derive gauge-invariant subleading operators and the time-ordered
products at a given order. Recent discussion \cite{bauer7,wise} on
the field redefinitions to make the gauge invariance explicit to all
orders, and the effects of the quark masses would help construct the
operators.

\section*{Acknowledgments}
We are grateful to Zoltan Ligeti for pointing out that the quark mass
effect on radiative $B$ decays can be of leading order in
$\mathrm{SCET}_{\mathrm{II}}$.  We add an appendix commenting on
the quark mass effects to complete the analysis. 
This work is supported by  R01-2002-000-00291-0 from the Basic
Research Program of KOSEF.

\appendix*
\section{Quark mass effects in radiative $B$ decays}
In Ref.~\cite{wise}, quark masses are included in the SCET
Lagrangian. If the strange quark mass scales as
$\Lambda_{\mathrm{QCD}}$, the strange quark mass terms are suppressed
in $\mathrm{SCET}_{\mathrm{I}}$, but are leading order in
$\mathrm{SCET}_{\mathrm{II}}$. The leading-order Lagrangian is given
as
\begin{equation}
\mathcal{L}_0 = \overline{\xi} \Bigl[in\cdot D
  +\Bigl(\FMSlash{\mathcal{P}}_{\perp} + g\FMSlash{A}_{n\perp} \Bigr) 
  W\frac{1}{\overline{n} \cdot \mathcal{P}} W^{\dagger}
  \Bigl(\FMSlash{\mathcal{P}}_{\perp} + g\FMSlash{A}_{n\perp} \Bigr)
  \Bigr]   \frac{\FMslash{\overline{n}}}{2} \xi,
\end{equation}
and the mass terms are given by
\begin{equation}
\mathcal{L}_m = m \overline{\xi}
\Bigl[\Bigl(\FMSlash{\mathcal{P}}_{\perp} + g\FMSlash{A}_{n\perp}
  \Bigr),W\frac{1}{\overline{n} \cdot \mathcal{P}} W^{\dagger}\Bigr]
\frac{\FMslash{\overline{n}}}{2} \xi -m^2 \overline{\xi}
W\frac{1}{\overline{n} \cdot \mathcal{P}} W^{\dagger}
\frac{\FMslash{\overline{n}}}{2} \xi.   
\end{equation}
We will consider only the case of the strange quark since the up and
the down quark masses are very small. As discussed in
Ref.~\cite{wise}, if the strange quark mass scales as
$\Lambda_{\mathrm{QCD}}$, $\mathcal{L}_m$ is suppressed. However, when
we go down to $\mathrm{SCET}_{\mathrm{II}}$, there is an enhancement
and we have to consider the operators to order $\lambda$ in
$\mathrm{SCET}_{\mathrm{I}}$ to obtain a leading-order result for
radiative $B$ decays. Therefore we include the effect of the first
term in $\mathcal{L}_m$. 

\begin{figure}[t]
\begin{center}
\epsfig{file=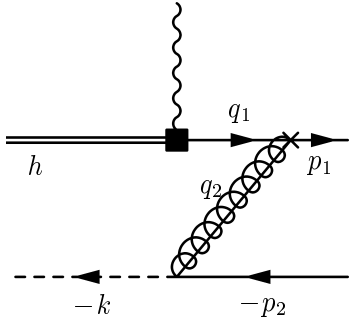, width=4.6cm}
\end{center}
\caption{Feynman diagram for the time-ordered product $T_m$ from
  $\mathcal{L}_m$. The square represents $O_7^{(0)}$ and the cross
  represents $\mathcal{L}_m$. The momentum $q_2$ of the gluon is
  incoming to $\mathcal{L}_m$.}
\label{fig11}
\end{figure}

The additional time-ordered product due to the mass terms at leading
order in $\mathrm{SCET}_{\mathrm{I}}$ is given by
\begin{equation}
T_m = \int d^4 x d^4 y T \Bigl[ O_7^{(0)} (0), i\mathcal{L}_m (x),
  i\mathcal{L}_{\xi q}^{(1)}(y) \Bigr].
\end{equation}
The Feynman diagram for $T_m$ is given by Fig.~\ref{fig10}.  The
time-ordered product in which there is no gluon from the vertex given
by $\mathcal{L}_m$ becomes zero when we match the operator to
$\mathrm{SCET}_{\mathrm{II}}$ because it is proportional to
$p_{\perp}^{\mu}$. The time-ordered product is written as
\begin{eqnarray}
T_m &=& \frac{em_b^2}{4\pi^2} g^2 m \frac{C_F}{N} \int \frac{d^4 x d^4
  q_1}{(2\pi)^4}   \frac{d^4 y d^4
  q_2}{(2\pi)^4}  e^{-iq_1 \cdot x-iq_2 \cdot (x-y)} e^{ip_1\cdot x}
  e^{-ik\cdot y} e^{ip_2\cdot y} \nonumber \\
&&\times \overline{\xi}_i  \Bigl( \gamma_{\perp\mu}
  \frac{1}{\overline{n} \cdot q_1} - \frac{1}{\overline{n} \cdot (q_1
  +q_2)} \gamma_{\perp \mu} \Bigr) \frac{\overline{n} \cdot
  q_1}{q_1^2} \FMSlash{\mathcal{A}} (1-\gamma_5)h_j \frac{1}{q_2^2}
  \overline{q}_{us,j} \gamma_{\perp}^{\mu} \xi_i, 
\end{eqnarray}
where the momenta are specified in Fig.~\ref{fig11}. Here $i$, $j$ are
color indices, and we omit the
Wilson lines, but after the calculation, we can always make the
operator gauge-invariant.

In going down to $\mathrm{SCET}_{\mathrm{II}}$, after we decouple soft 
gluons as was done in Eq.~(\ref{softd}), we can evaluate the matrix
element of the resultant operator. However, the basic form of the
operator to be evaluated is given by
\begin{eqnarray}
\overline{\xi}_i \gamma_{\perp\mu} \FMSlash{\mathcal{A}} (1-\gamma_5)
h_j \cdot \overline{q}_{s,j} \gamma_{\perp}^{\mu} \xi_i &=& \frac{1}{2}
\overline{\xi}_i \gamma_{\mu} (1+\gamma_5) \FMSlash{\mathcal{A}} h_j
\cdot \overline{q}_{s,j} \gamma^{\mu} (1-\gamma_5) \xi_i \nonumber \\
&&+\frac{1}{2} \overline{\xi}_i \gamma_{\mu} (1+\gamma_5)
\FMSlash{\mathcal{A}} h_j \cdot \overline{q}_{s,j} \gamma^{\mu}
(1+\gamma_5) \xi_i   \nonumber \\
&=& -\overline{\xi}_i (1-\gamma_5) \xi_i \cdot \overline{q}_{s,j}
(1+\gamma_5) \FMSlash{\mathcal{A}} h_j \nonumber \\
&&+\frac{1}{2} \overline{\xi}_i \gamma_{\mu}
(1+\gamma_5) \xi_i \cdot \overline{q}_{s,j} \gamma^{\mu} (1+\gamma_5)
\FMSlash{\mathcal{A}} h_j,
\label{massop}
\end{eqnarray}
where the Wilson lines and the momentum factors are omitted for
simplicity. When we 
take the matrix elements of the final operators in Eq.~(\ref{massop}),
they vanish at leading order in SCET. The quark mass effect may appear
at higher orders in $\alpha_s$ or at subleading order, but at the
order we consider, there is no quark mass effect.

\end{document}